\documentclass[zpreprint,zbstdefault]{./LaTeX/zeus/zeus_paper}
\usepackage[english]{babel}
\chardef\usc=95
\chardef\til=126
\catcode`\@=11 
\DeclareRobustCommand\xdotspace{\futurelet\@let@token\@xdotspace}
\def\@xdotspace{%
  \ifx\@let@token.\else
  \ifx\@let@token\bgroup.\else
  \ifx\@let@token\egroup.\else
  \ifx\@let@token\/.\else
  \ifx\@let@token\ .\else
  \ifx\@let@token~.\else
  \ifx\@let@token!.\else
  \ifx\@let@token,.\else
  \ifx\@let@token:.\else
  \ifx\@let@token;.\else
  \ifx\@let@token?.\else
  \ifx\@let@token/.\else
  \ifx\@let@token'.\else
  \ifx\@let@token).\else
  \ifx\@let@token-.\else
  \ifx\@let@token\@xobeysp.\else
  \ifx\@let@token\space.\else
  \ifx\@let@token\@sptoken.\else
   .\space
   \fi\fi\fi\fi\fi\fi\fi\fi\fi\fi\fi\fi\fi\fi\fi\fi\fi\fi}
\catcode`\@=12 

\newcommand{\stru}[2]{%
   \relax\ifmmode\hbox{\vrule height#1 depth#2 width0pt}%
   \else\vrule height#1 depth#2 width0pt\fi}

\newcommand{\Ronum}[1]{\uppercase\expandafter{\romannumeral#1}}
\newcommand{\ronum}[1]{\expandafter{\romannumeral#1}}
\DeclareRobustCommand{\LaTeXZ}{%
  \LaTeX\kern-.05em4\kern-.1em
  {\raisebox{-0.2ex}{$\scriptstyle\text{ZEUS}$}}\xspace}

\DeclareMathAlphabet{\mathbf}{OT1}{cmr}{bx}{sl}
\newcommand{\eVdist}{\kern-0.06667em}

\newcommand{\gev}{{\,\text{Ge}\eVdist\text{V\/}}}

\newcommand{\pb}{\,\text{pb}}

\newcommand{\pbi}{\,\text{pb}^{-1}}

\newcommand{\rad}{\,\text{rad}}

\newcommand{\slashfrac}[2]{%
  \raisebox{0.5ex}{\ensuremath #1}\kern-0.12em/\kern-0.08em
  \raisebox{-.8ex}{\ensuremath #2}}

\newcommand{\sqr}[3]{%
    {\vcenter{\hrule height.#3ex\hbox{\vrule width.#2ex height#1ex
     \kern#1ex\vrule width.#3ex}\hrule height.#2ex}}}

\catcode`\@=11 
\newcommand{\parenbar}{\mathpalette\p@renb@r}
\def\p@renb@r#1#2{\vbox{%
  \ifx#1\scriptscriptstyle \dimen@.7em\dimen@ii.2em\else
  \ifx#1\scriptstyle \dimen@.8em\dimen@ii.25em\else
  \dimen@1em\dimen@ii.4em\fi\fi \offinterlineskip
  \ialign{\hfill##\hfill\cr
    \vbox{\hrule width\dimen@ii}\cr
    \noalign{\vskip-.3ex}%
    \hbox to\dimen@{$\mathchar300\hfil\mathchar301$}\cr
    \noalign{\vskip-.3ex}%
    $#1#2$\cr}}}
\catcode`\@=12 

\newcommand{\IP}{{\rm I$\kern-0.01667em$P}\xspace}

\newcommand{\Lumi}{{\cal L}}

\mathchardef\qsm=63
\mathchardef\pls=43
\mathchardef\mns=512
\mathchardef\plm=518
\mathchardef\eql=61
\mathchardef\smallleft=300
\mathchardef\smallright=301
\mathchardef\les=316
\mathchardef\gre=318
\mathchardef\leq=532
\mathchardef\grq=533

\catcode`\@=11 
\newcounter{pict@width}
\newcounter{pict@height}
\newlength{\pict@scale}
\newcommand{\psfigadd}[4]{%
\setcounter{pict@width}{1*\ratio{#2+\pict@scale/2}{\pict@scale}}
\setcounter{pict@height}{1*\ratio{#3+\pict@scale/2}{\pict@scale}}
\setlength{\unitlength}{\pict@scale}
\hbox to #2{\hspace{-\fill}\begin{picture}(\thepict@width,\thepict@height)
\put(0,0){\psfig{figure=#1,width=#2,height=#3,clip=}}
\SetScale{0.283466457}
\SetWidth{1.763889}
{#4}
\end{picture}}
}
\newcounter{pict@widthfst}
\newcounter{pict@widthscd}
\newcounter{pict@widthtot}
\newcommand{\psfigaddtwo}[7]{%
\setcounter{pict@widthfst}{1*\ratio{#2+\pict@scale/2}{\pict@scale}}
\setcounter{pict@widthscd}{1*\ratio{#2+#4+\pict@scale/2}{\pict@scale}}
\setcounter{pict@widthtot}{1*\ratio{#2+#4+#6+\pict@scale/2}{\pict@scale}}
\setcounter{pict@height}{1*\ratio{#3+\pict@scale/2}{\pict@scale}}
\setlength{\unitlength}{\pict@scale}
\hbox{\hspace{-\fill}\begin{picture}(\thepict@widthtot,\thepict@height)
\put(0,0){\psfig{figure=#1,width=#2,height=#3,clip=}}
\put(\thepict@widthscd,0){\psfig{figure=#5,width=#6,height=#3,clip=}}
\SetScale{0.283466457}
\SetWidth{1.763889}
{#7}
\end{picture}}
}
\newcommand{\psfigror}[4]{%
\setcounter{pict@width}{1*\ratio{#2+\pict@scale/2}{\pict@scale}}
\setcounter{pict@height}{1*\ratio{#3+\pict@scale/2}{\pict@scale}}
\setlength{\unitlength}{\pict@scale}
\hbox{\begin{picture}(\thepict@width,\thepict@height)
\put(0,\thepict@height){\psfig{figure=#1,width=#3,height=#2,clip=,angle=270}}
\SetScale{0.283466457}
\SetWidth{1.763889}
{#4}
\end{picture}}
}
\newcommand{\psfigrol}[4]{%
\setcounter{pict@width}{1*\ratio{#2+\pict@scale/2}{\pict@scale}}
\setcounter{pict@height}{1*\ratio{#3+\pict@scale/2}{\pict@scale}}
\setlength{\unitlength}{\pict@scale}
\hbox{\begin{picture}(\thepict@width,\thepict@height)
\put(0,0){\psfig{figure=#1,width=#3,height=#2,clip=,angle=90}}
\SetScale{0.283466457}
\SetWidth{1.763889}
{#4}
\end{picture}}
}
\catcode`\@=12 
\newlength\listtextwidth

\newcommand{\pcite}[1]{{\protect\cite{#1}}}

\catcode`\@=11 
\newlength{\@tabfninsert}
\newlength{\@tabfnwidth}
\newcommand{\tabfootnote}[2]{%
  \setlength{\@tabfninsert}{0.8em}
  \setlength{\@tabfnwidth}{\textwidth}
  \addtolength{\@tabfnwidth}{-\@tabfninsert}
  \addtolength{\@tabfnwidth}{-0.4em}
  \noindent\makebox[\@tabfninsert][r]{\footnotesize$^{#1}$\hfil}\hfill%
  \parbox[t]{\@tabfnwidth}{\footnotesize #2\hfill}}
\catcode`\@=12 

\def\citeDjango{{\cite{%
cpc:81:381%
}}\xspace}
\def\citeHeracles{{\cite{%
cpc:69:155%
}}\xspace}
\def\citeLepto{{\cite{%
cpc:101:108%
}}\xspace}
\def\citeAriadne{{\cite{%
cpc:71:15%
}}\xspace}

\def\citeJetset{{\cite{%
cpc:39:347,*cpc:43:367,*cpc:82:74%
}}\xspace}
\def\citeEPVEC{{\cite{%
np:b375:3%
}}\xspace}

\def\citeGrape{{\cite{%
cpc:136:126%
}}\xspace}
\def\citePythia{{\cite{%
cpc:46:43,*cpc:82:74%
}}\xspace}
\def\citeHEXF{{\cite{%
lsuhe-145-1993%
}}\xspace}
\def\citeChinhan{{\cite{%
thesis:nguyen:2002%
}}\xspace}

\begin{document}
\prepnum{DESY--03--182}

\title{
Isolated tau leptons in events with large missing
transverse momentum at HERA\\
}                                                       
                    
\author{ZEUS Collaboration}
\draftversion{2.3 (post reading)}
\date{November 2003}

\abstract{
A search for events containing isolated tau leptons 
and large missing transverse momentum, not originating from the tau decay,
has been performed with
the ZEUS detector at the electron-proton collider HERA,
using $130~\pb^{-1}$ of integrated luminosity. 
A search was made for
isolated tracks coming from hadronic tau decays. 
Observables based on the internal jet structure were exploited
to discriminate between tau decays and quark- or gluon-induced 
jets. Three tau candidates were found, while $0.40^{+0.12}_{-0.13}$  
were expected from Standard Model processes, such as charged
current deep inelastic scattering and single $W^{\pm}$-boson production. 
To search for heavy-particle decays, 
a more restrictive selection was applied to isolate tau leptons
produced together with a hadronic final state with high transverse momentum.
Two candidate events survive, while $0.20\pm 0.05$ events are 
expected from Standard Model processes.
}

\makezeustitle

\def\3{\ss}                                                                                        
\pagenumbering{Roman}                                                                              


\begin{center}                                                                                     
{                      \Large  The ZEUS Collaboration              }                               
\end{center}                                                                                       
  S.~Chekanov,                                                                                     
  M.~Derrick,                                                                                      
  D.~Krakauer,                                                                                     
  J.H.~Loizides$^{   1}$,                                                                          
  S.~Magill,                                                                                       
  S.~Miglioranzi$^{   1}$,                                                                         
  B.~Musgrave,                                                                                     
  J.~Repond,                                                                                       
  R.~Yoshida\\                                                                                     
 {\it Argonne National Laboratory, Argonne, Illinois 60439-4815}, USA~$^{n}$                       
\par \filbreak                                                                                     
  M.C.K.~Mattingly \\                                                                              
 {\it Andrews University, Berrien Springs, Michigan 49104-0380}, USA
% ...
\newpage                                                                                           
$^{\    1}$ also affiliated with University College London, London, UK \\                          
% ...
                                                           %
                                                           %
\newpage   
                                                           %
                                                           %
\begin{tabular}[h]{rp{14cm}}                                                                       
$^{a}$ &  supported by the Natural Sciences and Engineering Research                               
          Council of Canada (NSERC) \\                                                             
$^{b}$ &  supported by the German Federal Ministry for Education and                               
          Research (BMBF), under contract numbers HZ1GUA 2, HZ1GUB 0, HZ1PDA 5, HZ1VFA 5\\         
$^{c}$ &  supported by the MINERVA Gesellschaft f\"ur Forschung GmbH, the                          
          Israel Science Foundation, the U.S.-Israel Binational Science                            
          Foundation and the Benozyio Center                                                       
          for High Energy Physics\\                                                                
$^{d}$ &  supported by the German-Israeli Foundation and the Israel Science                        
          Foundation\\                                                                             
$^{e}$ &  supported by the Italian National Institute for Nuclear Physics (INFN) \\                
$^{f}$ &  supported by the Japanese Ministry of Education, Culture,                                
          Sports, Science and Technology (MEXT) and its grants for                                 
          Scientific Research\\                                                                    
$^{g}$ &  supported by the Korean Ministry of Education and Korea Science                          
          and Engineering Foundation\\                                                             
$^{h}$ &  supported by the Netherlands Foundation for Research on Matter (FOM)\\                   
$^{i}$ &  supported by the Polish State Committee for Scientific Research,                         
          grant no. 620/E-77/SPB/DESY/P-03/DZ 117/2003-2005\\                                      
$^{j}$ &  partially supported by the German Federal Ministry for Education                         
          and Research (BMBF)\\                                                                    
$^{k}$ &  partly supported by the Russian Ministry of Industry, Science                            
          and Technology through its grant for Scientific Research on High                         
          Energy Physics\\                                                                         
$^{l}$ &  supported by the Spanish Ministry of Education and Science                               
          through funds provided by CICYT\\                                                        
$^{m}$ &  supported by the Particle Physics and Astronomy Research Council, UK\\                   
$^{n}$ &  supported by the US Department of Energy\\                                               
$^{o}$ &  supported by the US National Science Foundation\\                                        
$^{p}$ &  supported by the Polish State Committee for Scientific Research,                         
          grant no. 112/E-356/SPUB/DESY/P-03/DZ 116/2003-2005,2 P03B 13922\\                       
$^{q}$ &  supported by the Polish State Committee for Scientific Research,                         
          grant no. 115/E-343/SPUB-M/DESY/P-03/DZ 121/2001-2002, 2 P03B 07022\\                    
\end{tabular}                                                                                      
                                                           %

\pagenumbering{arabic} 
\pagestyle{plain}
\section{Introduction}
\label{sec-int}
Events with isolated leptons and large missing transverse momentum in
$e^\pm p$-collisions at HERA
can be a signature for processes beyond the Standard Model (SM).
The H1 and ZEUS Collaborations have previously reported searches for such events in
the cases where the lepton is an electron\footnote{Here and in the following, 
the term `electron' denotes generically both
the electron ($e^-$) and the positron ($e^+$).} or a muon
\cite{epj:c5:575,desy-02-224,pl:b471:411,desy-03-012}.
This paper presents a search for events with an isolated tau lepton and
missing transverse momentum which does not originate from the tau decay
($ep\rightarrow \tau \chi X$, where $\chi$ denotes one or more particles not interacting
inside the detector).
Such events are expected to occur at low rates in the SM from
decays of $W^\pm$ bosons into $\tau^\pm \parenbar{\nu_\tau}$, where the $W^\pm$ is produced radiatively
from the quark or the beam lepton.
Events with a large hadronic transverse momentum in addition to an isolated lepton are 
of particular interest since the SM background falls steeply with increasing hadronic
transverse momentum. Such events may result from the decay of a heavy particle.
One possible source for this signature would be the production of
single top quarks through flavour changing neutral currents (FCNC),  
with subsequent decay $t \rightarrow bW^+$, as predicted
by many theories beyond the SM
\cite{np:b454:527,*pl:b426:393,*pr:d58:073008,*pr:d60:074015,*pl:b457:186}.
Production of stop quarks in $R$-parity ($R_p$) violating SUSY 
models~\cite{np:b397:3,*pl:b270:81,*pr:d59:095009}
with subsequent two-body decay (e.g. $\tilde{t} \rightarrow \tau b$) or
$R_p$-conserving three-body decay modes
($\tilde{t}\rightarrow \tau \tilde{\nu_\tau}b,~\tilde{\tau} \nu_\tau b$)
are also potential sources.

The tau leptons were identified from their hadronic decay by requiring a 
collimated and low-multiplicity hadronic jet.
Charged current (CC) and neutral current (NC) interactions, with
gluon- and quark-induced jets, are large potential backgrounds to this process.
Restrictive conditions applied to jets reduced such backgrounds to 
a rate comparable to that of single $W^\pm$ production.

This paper is organized as follows. Section \ref{sec-det} describes the ZEUS
detector and the experimental conditions. Section \ref{sec-simulation} introduces the 
$e^\pm p$-interaction processes that were considered in this analysis, and their Monte 
Carlo simulation. 
The identification of tau leptons, which 
is based on an independent study, is introduced in Section \ref{sec-tau}. 
Section \ref{sec-analysis} presents the selection requirements for events with isolated 
tau leptons. The results of the analysis are discussed in Section \ref{sec-results}.
Section \ref{sec-conclusion} gives the conclusions.

\section{Experimental conditions}
\label{sec-det}

The data used in this analysis were collected with the ZEUS detector 
at HERA and correspond to an integrated luminosity of
$47.9\pm 0.9\ (65.5\pm 1.5)\pbi$ for $e^+p$ collisions taken during
1994-1997 (1999-2000) and $16.7\pm 0.3\pbi$ for $e^-p$ collisions
taken during {1998-99}. During {1994-97} ({1998-2000}), HERA operated with
protons of energy $E_p=820$~GeV ($920$~GeV) and electrons
of energy $E_e=27.5$~GeV, yielding a centre-of-mass energy of
$\sqrt s=300$~GeV ($318$~GeV).

The ZEUS detector is described in detail
elsewhere~\cite{zeus:1993:bluebook,pl:b293:465}. The main components
used in this analysis were the central tracking detector
(CTD)~\cite{nim:a279:290,*npps:b32:181,*nim:a338:254}, positioned in a
1.43~T solenoidal magnetic field, and the uranium-scintillator sampling
calorimeter (CAL)~\cite{nim:a309:77,*nim:a309:101,*nim:a321:356,*nim:a336:23}.

Tracking information is provided by the CTD, in which the momenta of
tracks in the polar-angle\footnote{The ZEUS coordinate system is a
right-handed Cartesian system, with the $Z$ axis pointing in the proton
beam direction, referred to as the ``forward direction'', and the $X$
axis pointing left towards the centre of HERA. The coordinate origin is
at the nominal interaction point.} region $15^\circ < \theta < 164^\circ$
are reconstructed. The CTD consists of 72 cylindrical drift chamber
layers, organised in nine superlayers. The relative transverse-momentum
resolution for full-length tracks can be parameterised as
$\sigma(p_T)/p_T=0.0058\ p_T \oplus 0.0065\oplus 0.0014/p_T$, with
$p_T$ in GeV.

The CAL covers $99.7\%$ of the total solid angle. It is divided into
three parts with a corresponding division in $\theta$, as viewed
from the nominal interaction point: forward (FCAL,
$2.6^{\circ}<\theta<36.7^{\circ}$), barrel (BCAL,
$36.7^{\circ}<\theta<129.1^{\circ}$), and rear (RCAL,
$129.1^{\circ}<\theta<176.2^{\circ}$). Each of the CAL parts
is subdivided into towers which in turn are segmented longitudinally into
one electromagnetic (EMC) and one (RCAL) or two (FCAL, BCAL) hadronic
(HAC) sections. The smallest subdivision of the CAL is called a cell.
Under test-beam conditions, the CAL single-particle relative energy 
resolution is
$\sigma(E)/E=0.18/\sqrt{E}$ for electrons
and $\sigma(E)/E=0.35/\sqrt{E}$ for hadrons, with $E$ in GeV.
In addition, the readout of the individual CAL cells provides timing information,
with a resolution better than 1~ns for energy depositions larger 
than $4.5\gev$.

The luminosity was measured using the Bethe-Heitler reaction
$e^\pm p\rightarrow e^\pm \gamma p$. The resulting small-angle energetic photons
were measured by the luminosity
monitor~\cite{desy-92-066,*zfp:c63:391,*acpp:b32:2025}, a lead-scintillator
calorimeter placed in the HERA tunnel at $Z=-107$~m.
A three-level trigger was used to select events online 
\cite{zeus:1993:bluebook,proc:chep:1992:222}.

\section{Monte Carlo simulation}
\label{sec-simulation}
In the following, processes which may lead to the event topology of
interest, and their Monte Carlo (MC) simulations, are described. All generated MC events
were passed through the GEANT 3.13-based \cite{tech:cern-dd-ee-84-1} ZEUS detector- and 
trigger-simulation programs \cite{zeus:1993:bluebook}. They were reconstructed and 
analysed by the same program chain as the data.

\emph{$W^\pm$ production}: $e^\pm p\rightarrow e^\pm W X$.
The production of real $W^\pm$ bosons with subsequent decay 
$W^\pm \rightarrow \tau^\pm \parenbar{\nu_\tau}$ is the only SM process with sizeable cross section leading to 
events with an isolated tau lepton and missing transverse momentum. 
Single $W^\pm$ production was simulated using the event generator EPVEC~\citeEPVEC.
The hadronisation of the partonic final state and the decays of the tau leptons were
performed by JETSET \citeJetset. As a cross check, control MC samples were used with
the tau decays performed by TAUOLA~2.6~\cite{tauola:91}.
Recent cross-section
calculations including $\mathcal{O}(\alpha^2 \alpha_s)$ QCD 
corrections~\cite{jp:g25:1434,*hep-ph/9905469,*hep-ph/0203269,*hep-ph/w_nlo_mc} 
and using the CTEQ4M~\cite{pr:d55:1280} (ACFGP~\cite{zfp:c56:589}) proton (photon) parton density
functions
were used to reweight the EPVEC samples. 
The total cross section for $W^\pm$ production is 1.0~pb (1.2~pb) for
an $e^\pm p$ centre-of-mass energy of $\sqrt{s}=300\gev$ ($318\gev$).
The contribution of the CC process $e^\pm p\rightarrow \parenbar{\nu_e} W^\pm X$
is about 5\% of that from the neutral current process and 
was neglected \cite{pl:b471:411}. 

\emph{Charged current deep inelastic scattering (CC DIS)}: $e^\pm p\rightarrow \parenbar{\nu_e} X$.
Events from CC DIS interactions can mimic the selected
topology if a particle from the hadronic final state is misidentified as
an isolated tau lepton. The CC DIS events were
simulated using the event generator DJANGO6 \citeDjango, an interface
to the MC programs HERACLES 4.5 \citeHeracles and LEPTO
6.5 \citeLepto. Leading-order QCD and electroweak radiative
corrections were included and higher-order QCD effects were simulated
via parton cascades using the colour-dipole model (CDM) as implemented in ARIADNE
\citeAriadne or matrix elements and parton showers (MEPS) based on a leading-logarithmic
approximation as implemented in LEPTO. The hadronisation of the partonic
final state was performed by JETSET. The CTEQ4D~\cite{pr:d55:1280} 
parameterisations for the parton density functions (PDFs) in the proton
were used.

\emph{Neutral current deep inelastic scattering (NC DIS)}: $e^\pm p\rightarrow e^\pm X$.
The scattered electron or a jet from the hadronic system in an NC DIS event 
can be misidentified as an isolated tau lepton. This can lead to the
selected event topology, if combined with apparent missing transverse momentum, which may 
arise from leptonic decays of charm or bottom quarks,
fluctuations in the detector response or
undetected particles due to the limited geometric acceptance of the
detector. The NC DIS events
were simulated in the same framework as the CC DIS events. 
The CTEQ5D~\cite{epj:c12:375} 
parameterisations for the proton PDFs were used.

\emph{Photoproduction of jets}: $\gamma p \rightarrow X$.
Background from hard scattering photoproduction processes can contribute
to the selected event topology if a particle from the hadronic final state
is misidentified as a tau lepton and apparent missing transverse momentum is present,
arising from the sources described above.
Resolved and direct photoproduction
processes were simulated using PYTHIA 5.7 \citePythia.

\emph{Lepton-pair production}: $e^\pm p\rightarrow e^\pm l^\pm l^\mp X$,~$l=e,\mu,\tau$.
Pair production of leptons via the Bethe-Heitler process can lead to 
events with the selected topology, if one of the leptons escapes 
detection or is misidentified as a QCD jet and a mismeasurement causes missing 
transverse momentum. Lepton-pair production was simulated using the GRAPE
dilepton generator~\citeGrape, including both the elastic and inelastic 
components at the proton vertex.

\emph{Single-top production in theories beyond the SM}: $e^\pm q\rightarrow e^\pm t \rightarrow e^\pm b W^+$.
A significant number of single top quarks with 
subsequent decays into a $b$ quark and a $W^+$ boson could be produced if the 
top quarks were to be produced via anomalous
effective couplings, including FCNC of the type 
$tuV$ $(u=$ up-type quark in the proton, $V=\gamma, Z^0$) 
\cite{np:b454:527,*pl:b426:393,*pr:d58:073008,*pr:d60:074015,*pl:b457:186}. 
An isolated tau lepton and a neutrino from the $W^+$ decay lead to the 
selected event topology. The large mass of the top quark could result in large
transverse momenta of its decay products, which through the subsequent
$b$-quark decay would produce a large hadronic transverse momentum in the detector. 
In the current paper, the anomalous production of single top quarks was used as a template for
processes involving the production of heavy particles with tau leptons in the 
decay chain.
Single-top production through FCNC processes in $e^\pm p$ collisions was
simulated using the HEXF generator~\citeHEXF. 

\section{Tau identification}
\label{sec-tau}
The search for tau leptons is based on their hadronic decays.
The narrow, ``pencil-like'',
shape and the low charged-particle multiplicity of the tau jets were
used to distinguish them from quark- and gluon-induced jets
\citeChinhan. 
\subsection{Jet observables}
\label{sec-jetobs}
The longitudinally
invariant $k_T$ cluster algorithm~\cite{np:b406:187} was used in the
inclusive mode~\cite{pr:d48:3160a} to reconstruct jets from the
energy deposits in the CAL cells. The jet search was performed in
the $\eta - \phi$ plane of the laboratory frame, where $\eta_i$ and $\phi_i$,
the pseudorapidity and azimuthal angle of each CAL cell, were calculated
using the primary event vertex as reconstructed in the CTD.
The axis of each jet
was defined according to the Snowmass
convention~\cite{proc:snowmass:1990:134}, where $\eta_\mathrm{jet}$
($\phi_\mathrm{jet}$) was the transverse-energy-weighted mean pseudorapidity
(azimuth angle) of all the cells belonging to the jet. The jet transverse
energy, $E_T^\mathrm{jet}$, was reconstructed as the sum of the transverse
energies of the cells belonging to the jet and was corrected for
detector effects such as energy losses in the inactive material in
front of the CAL~\cite{pl:b531:9}.

The internal jet structure is generally well described by the MC simulations
\cite{np:b545:3,desy-02-217}. For this analysis, it was characterised by 
six observables:

\begin{itemize}

  \item the first moment of the radial extension of the jet
    $$ R_\mathrm{mean} = \langle R \rangle = \frac{\sum_{i}E_i \cdot
    R_i}{\sum_{i}{E_i}},$$
    where the sum runs over the CAL cells associated to the jet,
    $E_i$ is the
    energy of the cell $i$ and $R_i$ is defined
    as $R_i=\sqrt{\Delta \phi_i^2 + \Delta \eta_i^2}$, where
    $\Delta \phi_i$ ($\Delta \eta_i$) is the difference between the
    azimuthal angle (pseudorapidity) of the calorimeter cell $i$ and
    the jet axis;

  \item the second moment of the
    radial extension of the jet
    $$ R_\mathrm{rms} =\sqrt{\frac{\sum_{i} E_i  (\langle R \rangle -  R_i)^2}{ \sum_{i}{E_i}}};$$
  \item the first moment of the projection of the jet onto its axis
    $$ L_\mathrm{mean} =  \langle L   \rangle =  \frac{\sum_{i}
    E_i \cdot \cos{\alpha_i}}{\sum_{i}E_i},$$
    where $\alpha_i$ 
    is the angle between the cell $i$ and the jet axis;
  \item the second moment of the projection of the jet onto its axis
    $$ L_\mathrm{rms} =  \sqrt{\frac{\sum_i{E_i(\langle L \rangle - \cos{\alpha_i})^2}}{\sum_i E_i}};$$

  \item the number of subjets ($N_\mathrm{subj}$) with a $y_{\rm{cut}}$ of $5 \cdot
    10^{-4}$. \\
    The subjet multiplicity identifies the number of localised energy depositions within a jet
    that can be resolved using a resolution-criterion $y_{\rm{cut}}$.
    The number of subjets was found by applying the same algorithm as
    was initially used to find jets. An exact definition can be found elsewhere 
    \cite{jhep:09:009,np:b421:545,desy-02-217}; 

   \item the invariant mass, $M_{\rm jet}$, of the jet four-vector, calculated from
     the cells associated to the jet. The particles of the jet were assumed to 
     be massless.
    
\end{itemize}

\subsection{Control selection}
The tau-identification procedure was determined using event samples selected independently 
from those used for the analysis. 
Monte Carlo events from single $W^\pm$ production, where the $W^\pm$ decays to a tau lepton
and a neutrino and the tau lepton decays hadronically ($W^\pm\rightarrow \tau^\pm \parenbar{\nu_\tau}, \tau \rightarrow
hadrons$), were 
used as signal. The background simulation was based on an inclusive 
sample of MC CC DIS events. 

An inclusive CC DIS data sample was used
to monitor the quality of the simulation~\cite{thesis:dannheim:2003}.
To obtain this sample, large missing transverse 
momentum and the existence of at least one jet with $E_T^\mathrm{jet}>5\gev$ 
in the polar-angle range $15^\circ<\theta_\mathrm{jet}<164^\circ$ were required. 
Electrons
from badly reconstructed NC DIS events with large apparent missing transverse momentum 
were suppressed by rejecting jets that were back-to-back 
with the hadronic system.
Remaining electrons were rejected based on the fraction 
of electromagnetic energy and on the fraction of the jet energy carried by the leading track 
pointing in the direction of the jet \cite{pr:d65:092004}. 

\begin{figure}[p]
\begin{center}
\begin{minipage}[t]{1.0\linewidth}
\begin{center}
\epsfig{figure=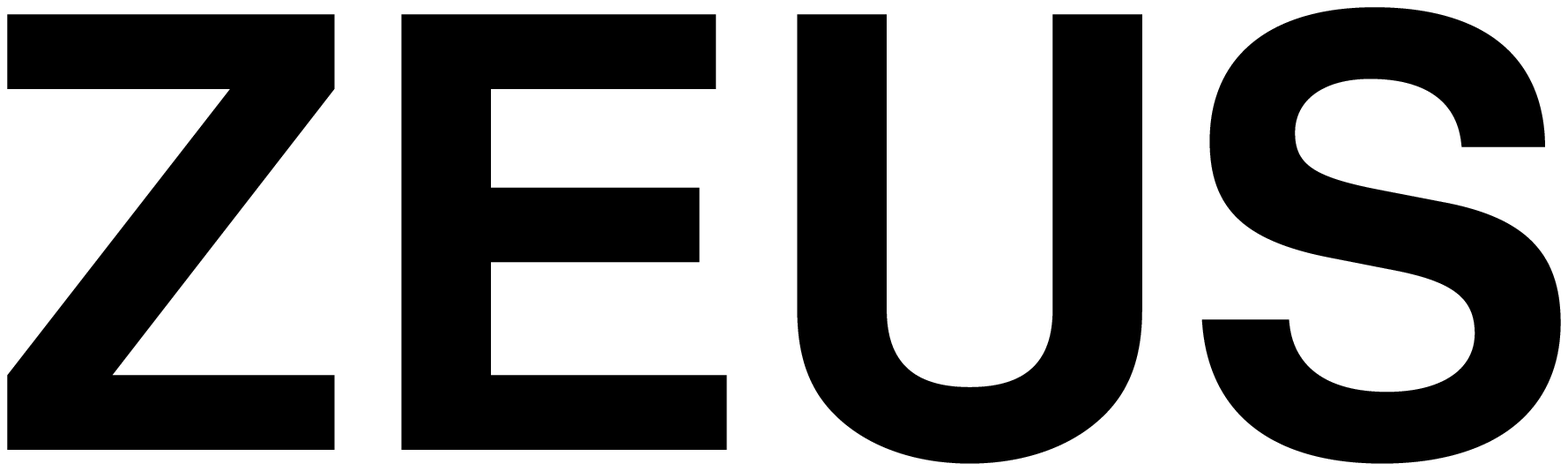, width={0.13\linewidth},angle=0,clip=}
\end{center}
\end{minipage}
\begin{minipage}[t]{0.5\linewidth}
\epsfig{figure = 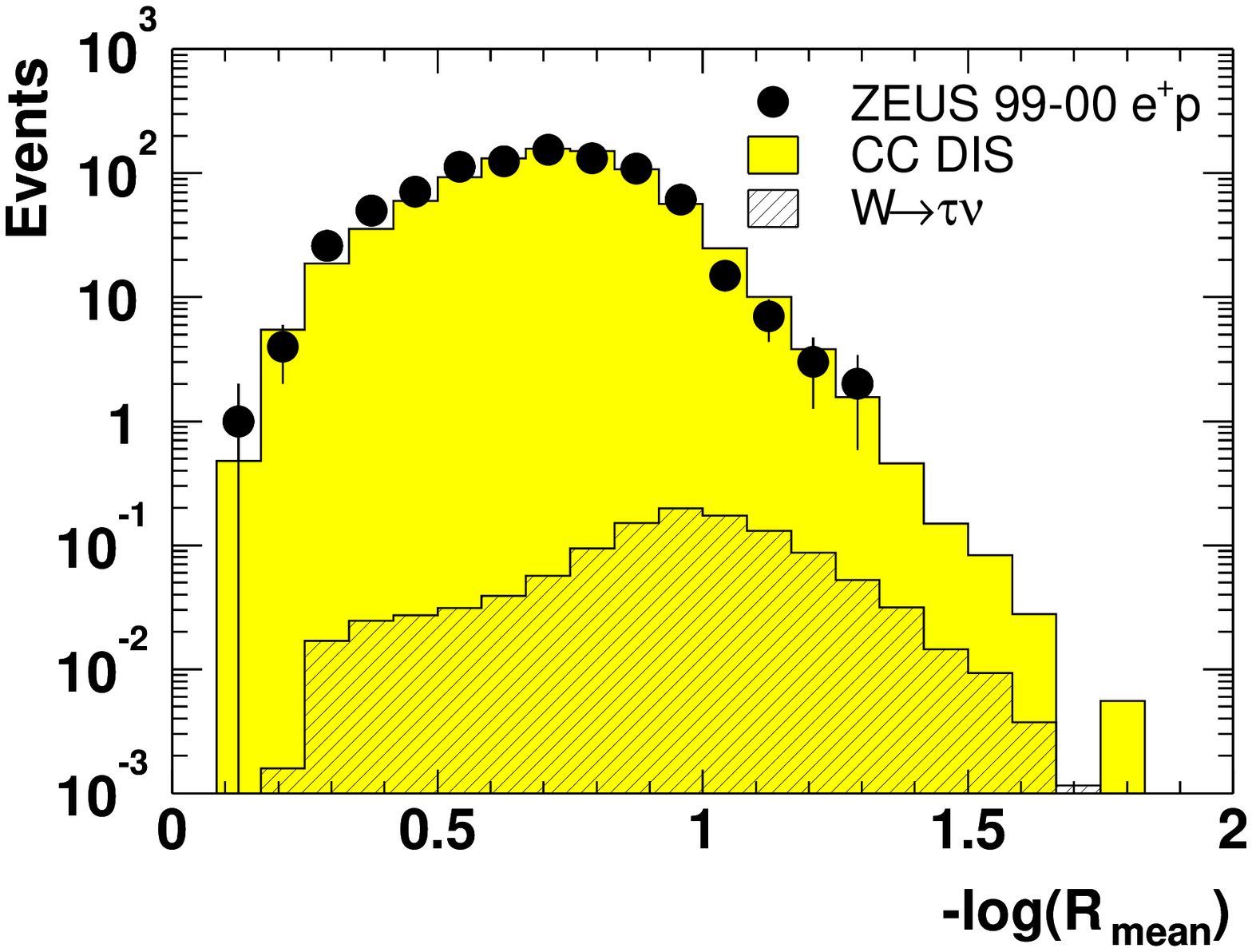,
width={1.0\linewidth}, angle=0, clip=}
\epsfig{figure = 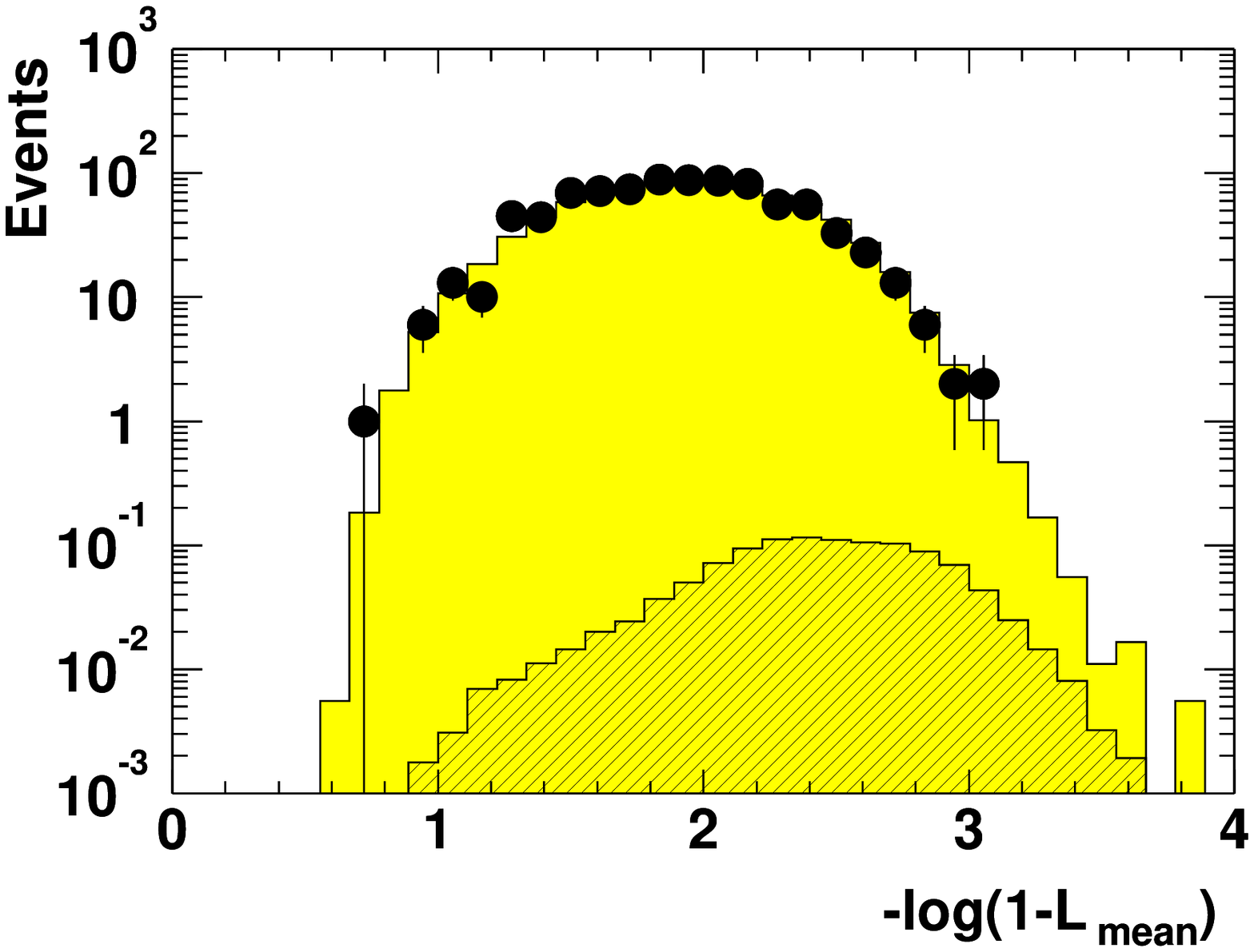,
width={1.0\linewidth}, angle=0, clip=}
\epsfig{figure = 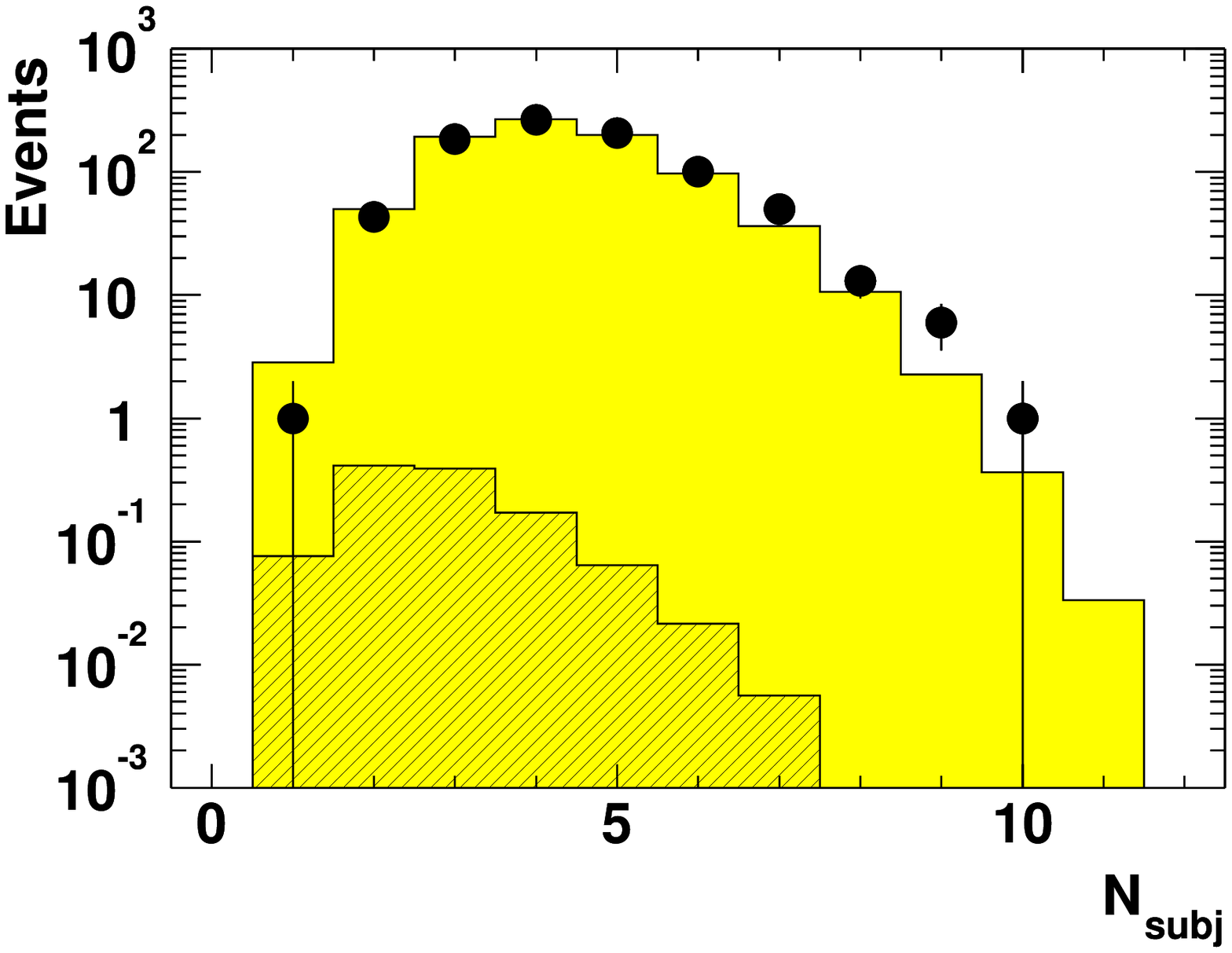,
width={1.0\linewidth}, angle=0, clip=}
\end{minipage}\hfill
\begin{minipage}[t]{0.5\linewidth}
\epsfig{figure = 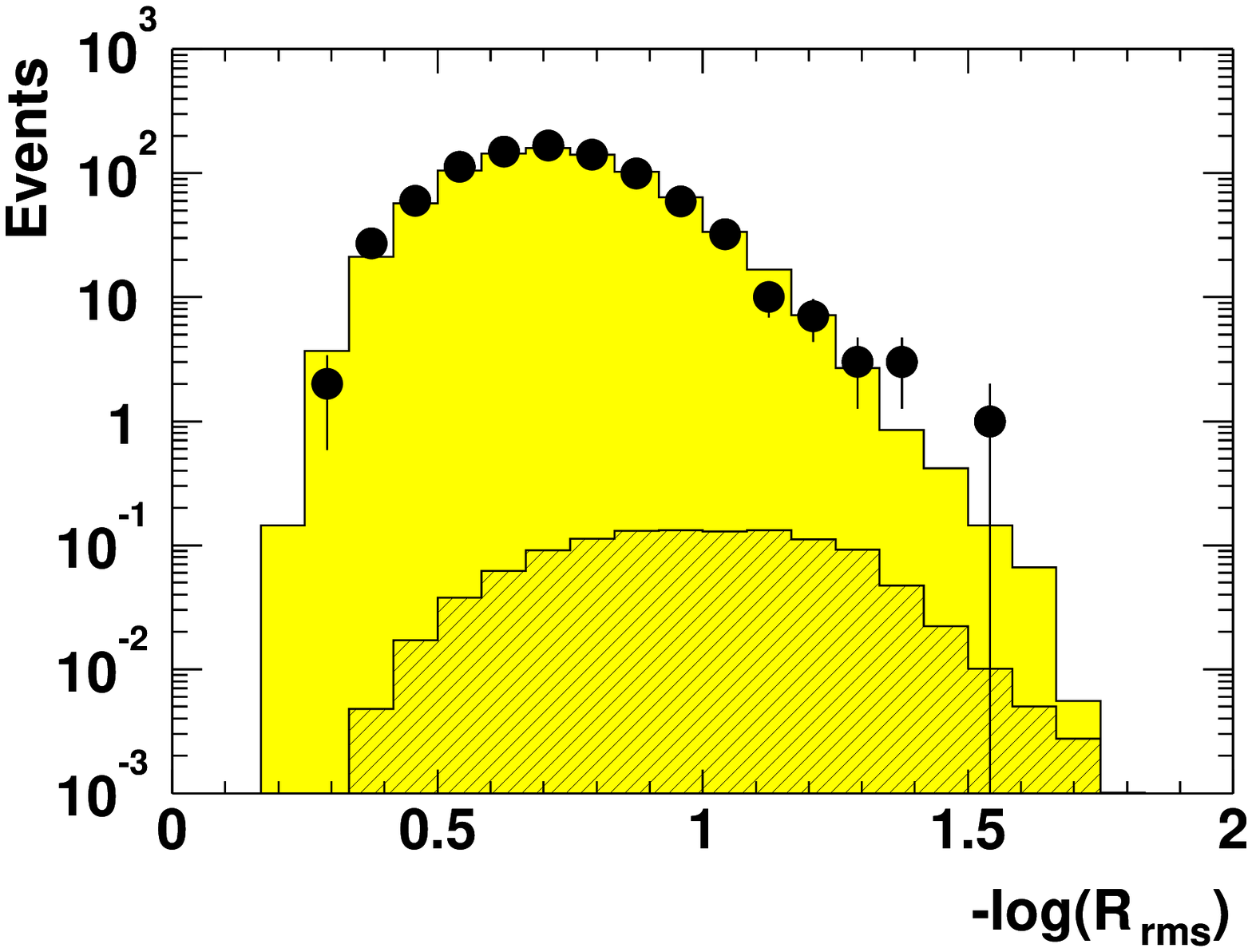,
width={1.0\linewidth}, angle=0, clip=}
\epsfig{figure = 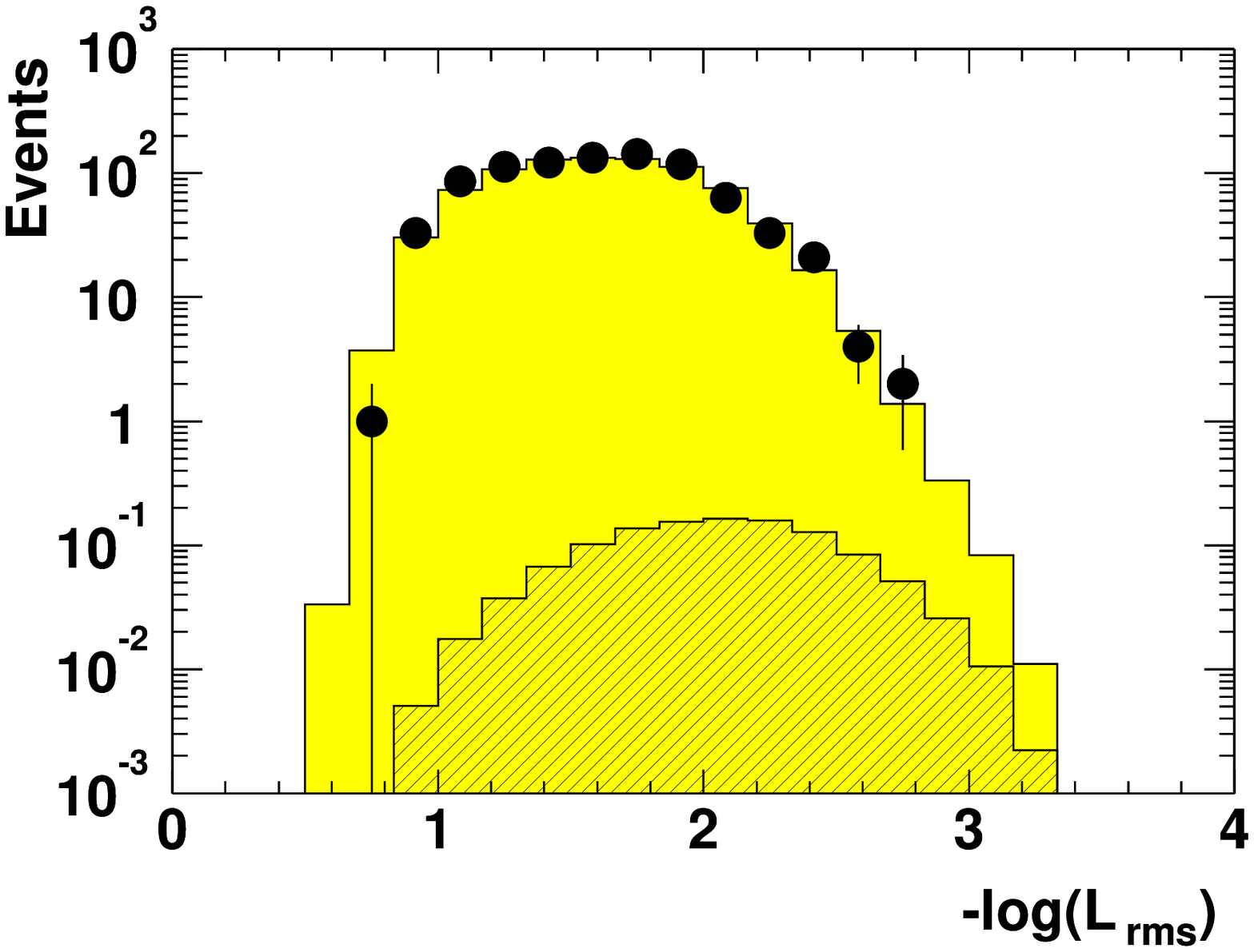,
width={1.0\linewidth}, angle=0, clip=}
\epsfig{figure = 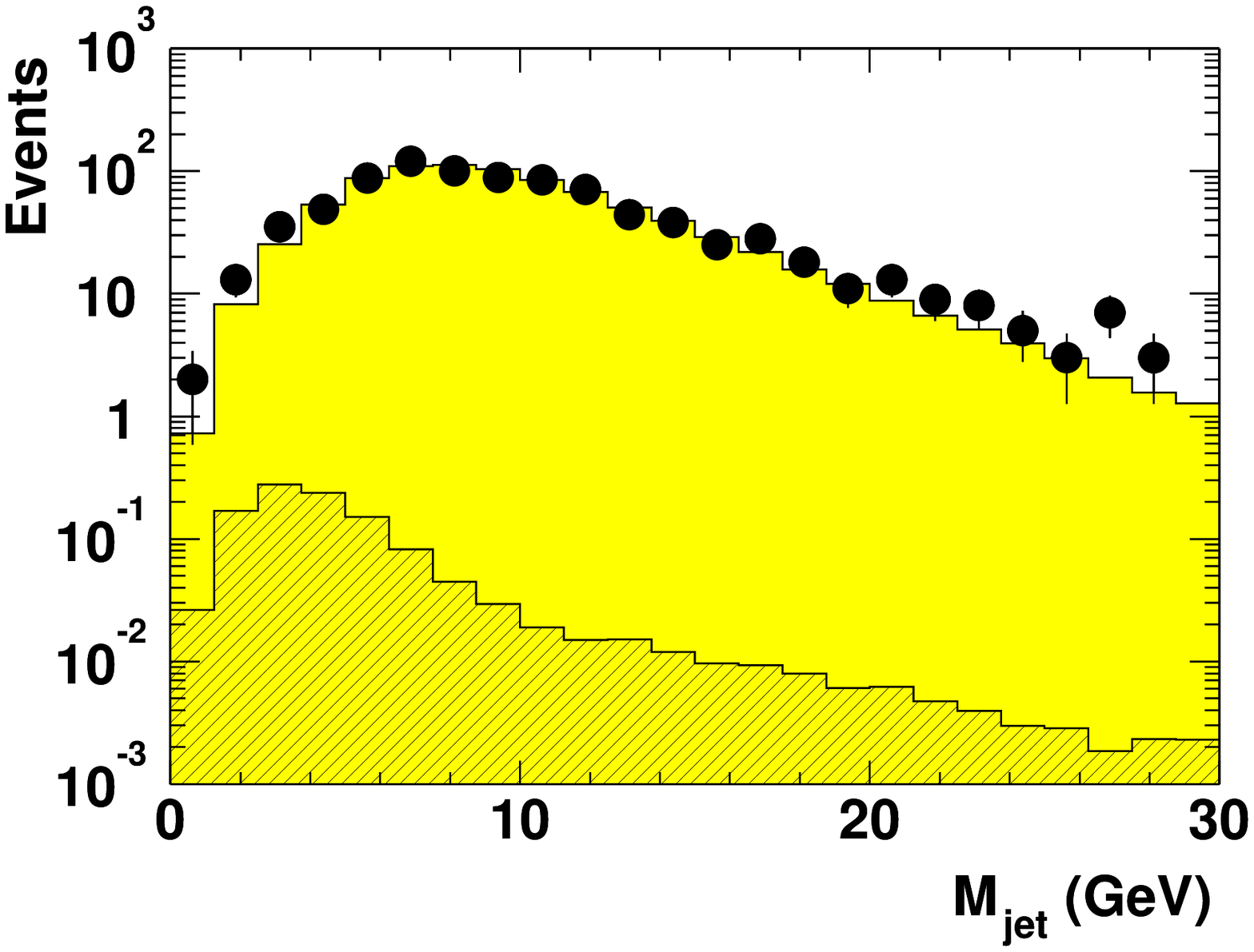,
width={1.0\linewidth}, angle=0, clip=}
\end{minipage}\hfill
\end{center}
\begin{picture} (0.,0.)
\setlength{\unitlength}{1.0cm}
\put (1.7,17.05){\bf\small (a)}
\put (9.7,17.05){\bf\small (b)}
\put (1.7,11.05){\bf\small (c)}
\put (9.7,11.05){\bf\small (d)}
\put (1.7,5){\bf\small (e)}
\put (9.7,5){\bf\small (f)}
\end{picture}
\caption{Observables characterising the internal jet structure for an
inclusive selection of CC DIS events (see text for definitions). Shown
are the data (dots), the simulation of CC DIS events (shaded histograms) 
and the simulation of
the direct $W^\pm$-production signal $W^\pm \rightarrow \protect\parenbar{\nu_\tau} \tau^\pm$, 
where the $\tau$ decays hadronically (hatched histograms).}
\label{fig-tauobs}
\end{figure}

Figure \ref{fig-tauobs} shows the comparison of the inclusive CC DIS data
sample and the MC CC events in each of the six jet-shape observables. For each event, only the
jet with the largest value of the tau discriminant, as defined below, enters.
The agreement between the data and the simulation is good. The expected signal from 
tau decay is also shown. A difference in the shapes between the tau jets and the quark- or
gluon-induced jets is evident for all six variables.

\subsection{Tau discriminant}
\begin{figure}[htbp]
\begin{minipage}[t]{1.0\linewidth}
\begin{center}
\epsfig{figure=plots/zeus.eps, width={0.13\linewidth},angle=0,clip=}
\end{center}
\end{minipage}
\begin{center} 
\epsfig{figure = 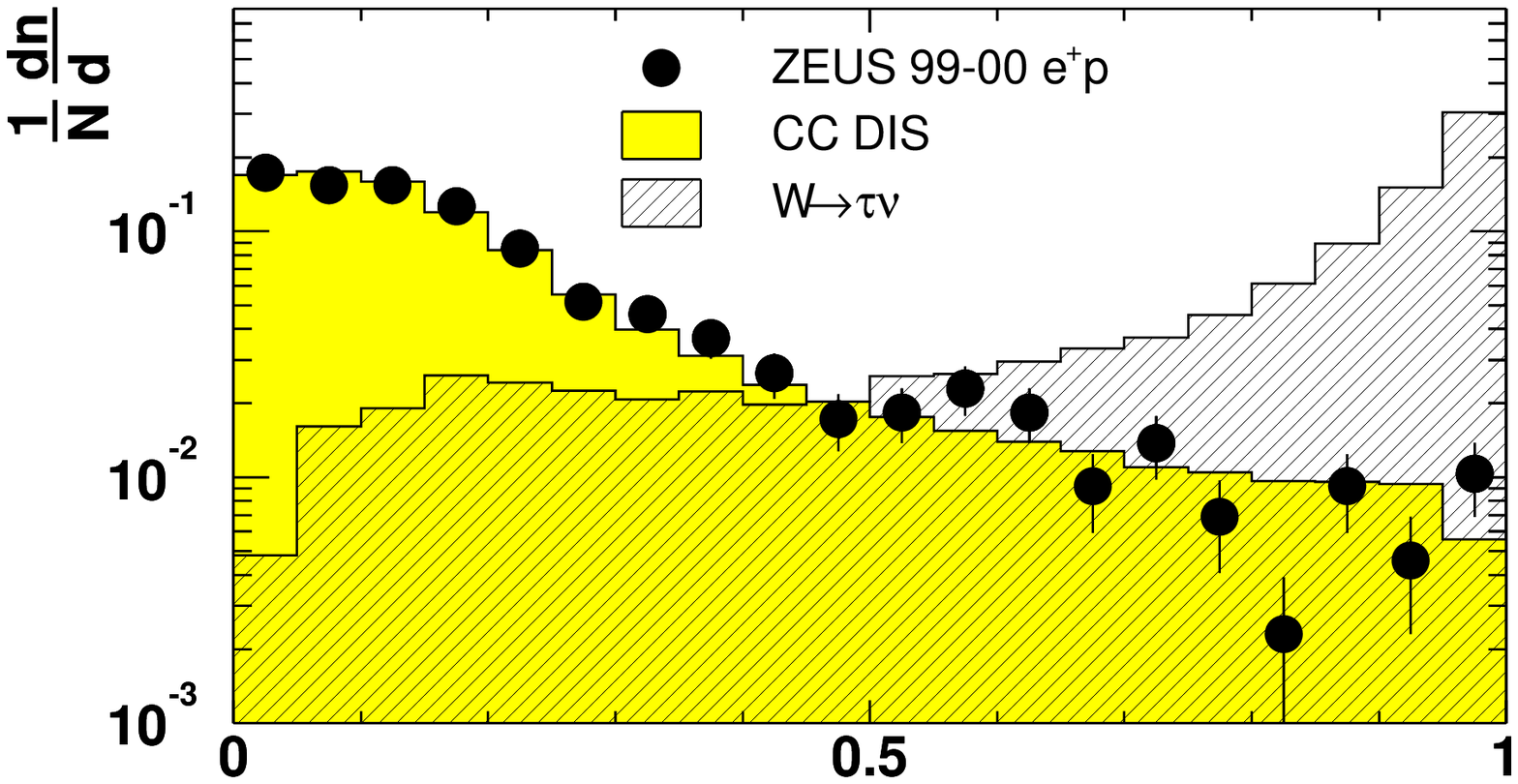,
width={14cm}, angle=0, clip=} \end{center}
\begin{center} \epsfig{figure = 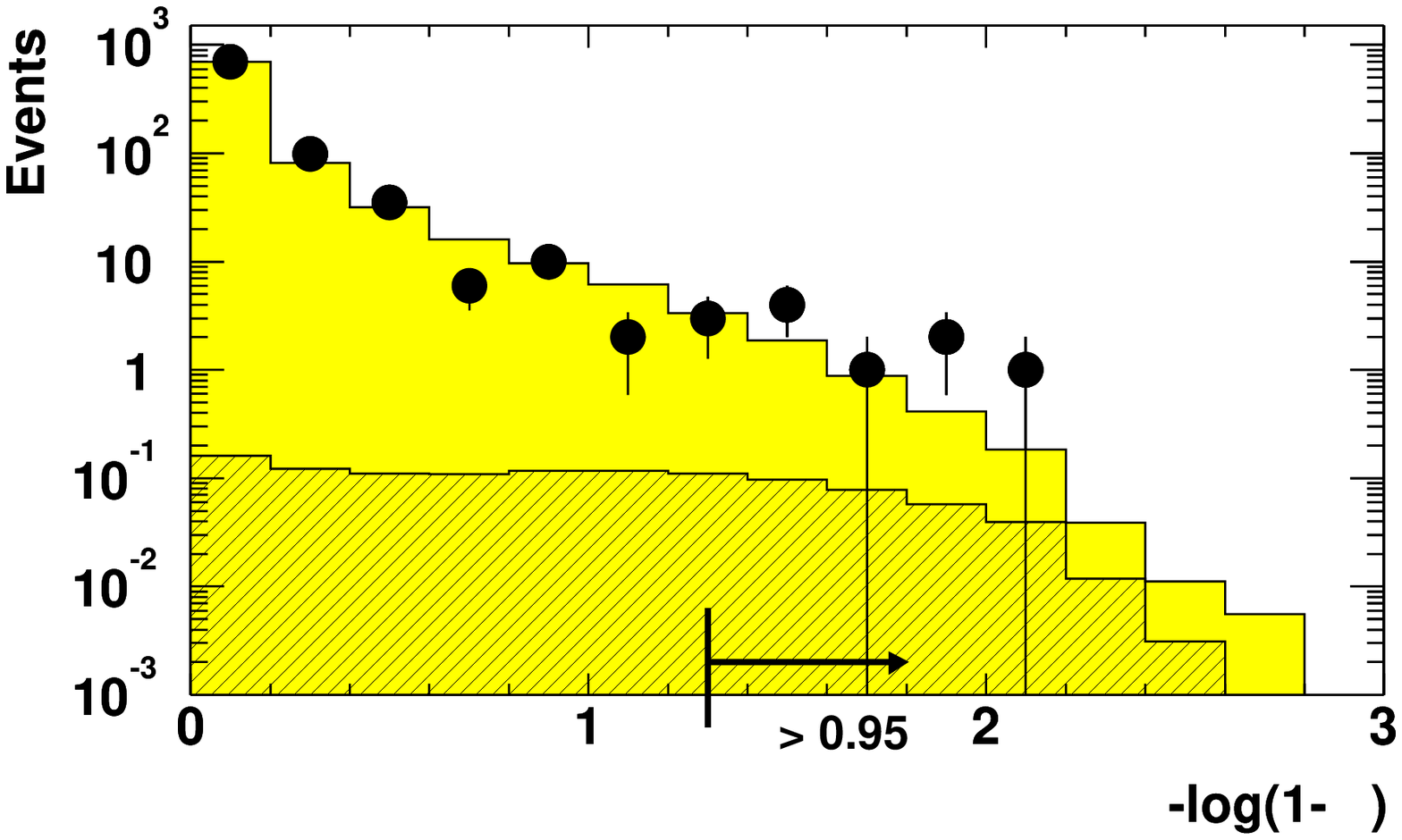,
width={14cm}, angle=0, clip=} \end{center}
\begin{picture} (0.,0.)
\setlength{\unitlength}{1.0cm}
\put (13.3,16.15){\bf\small (a)}
\put (13.3,7.7){\bf\small (b)}
\put (13.9,9.5){\large $\mathcal{D}$}
\put (8.33,1.628){\large $\mathcal{D}$}
\put (13.74,0.975){\large $\mathcal{D}$}
\begin{sideways}\put (16.3,-2.45){$\mathcal{D}$}\end{sideways}
\end{picture}
\caption{Distribution of the tau discriminant, $\mathcal{D}$, for 
an inclusive selection of CC DIS data events (dots), a 
simulation of CC DIS events (shaded histograms) and the 
simulation of the direct $W^\pm$-production
signal $W^\pm \rightarrow \protect\parenbar{\nu_\tau} \tau^\pm$, where 
the $\tau$ decays hadronically (hatched histograms). In each
event, only the jet with the highest value of the discriminant enters.
The histograms are normalised (a) to the total number of events N and
(b) to the luminosity of the data. 
In (b), the $-\log (1-\mathcal{D})$ distribution is displayed 
to expand the region in which the tau lepton signal is expected.}
\label{fig-taudiscr}
\end{figure}

To separate the signal from the background,
the six jet-shape observables were combined in a discriminant $\mathcal{D}$,
given for any point, $\vec x$,  in the phase space, where 
$$\vec x =
(\rm{-log(R_\mathrm{mean}), -\log(R_\mathrm{rms}), -\log(1-L_\mathrm{mean}), -\log(L_\mathrm{rms}), N_{subj}, M_{jet}}),$$
as:
$$\mathcal{D}(\vec x)=\frac{\rho_{\rm{sig}}(\vec x)}{\rho_{\rm{sig}}(\vec
x)+\rho_{\rm{bg}}(\vec x)},$$
where $\rho_{\rm{sig}}$ and $\rho_{\rm{bg}}$ are the density functions of the signal and
the background events, respectively.
The signal and background densities, sampled using MC simulations, were
calculated using a probability-density-estimation method based on range 
searching (PDE-RS) 
\cite{carli:modt:0011224,*carli:modt:0211019}. For any given jet with phase-space 
coordinates $\vec x$, the
signal and background densities were evaluated from the number of corresponding
signal and background jets in a six-dimensional box of fixed size 
centred around $\vec x$.
Figure~\ref{fig-taudiscr} shows the distribution of $\mathcal{D}$
for the MC-generated signal and background events and for the data selection. 
For each event, the jet with the largest value of the discriminant enters.
The data are
well described by the MC simulation for the inclusive CC selection. 
The tau signal tends to have large discriminant values ($\mathcal{D}\rightarrow1$) and
is clearly separated from the CC DIS background at low discriminant values ($\mathcal{D}\rightarrow 0$). 

The quality of the tau selection is 
characterised by the efficiency of the signal selection $\epsilon _\mathrm{sig}$, the rejection of 
the background, $R$, and the separation power, $S$, which are 
defined for a given cut on the discriminant, $\mathcal{D}_\mathrm{cut}$, as follows:
\begin{eqnarray}
\label{eqn-effrejsep}
\epsilon_{\mathrm{sig}}&=&N_{\mathrm{sig,selected}}/N_{\mathrm{sig,total}} \nonumber \\
R&=&N_{\mathrm{bg,total}}/N_{\mathrm{bg,selected}} \nonumber \\
S&=&\sqrt{R}\cdot \epsilon_{\mathrm{sig}}. \nonumber
\end{eqnarray}
In the equations above,
$N_{\mathrm{sig,total}}$ and $N_{\mathrm{bg,total}}$ are the total number of signal and background
events, respectively, and $N_{\mathrm{sig,selected}}$ and $N_{\mathrm{bg,selected}}$ are the number of
signal and background events after applying a cut of $\mathcal{D}>\mathcal{D}_\mathrm{cut}$, respectively.
The cut on $\mathcal{D}$ was optimised for maximal separation power. The optimisation resulted in a value of
$\mathcal{D}>0.95$, for which a signal efficiency $\epsilon _{\mathrm{sig}}=31\pm0.2\%$, a background
rejection $R=179\pm6$ and a separation power $S=4.1\pm0.1$ were obtained. 
The quoted uncertainties are the statistical uncertainties due to the limited number of generated
MC events.
When restricting the selection to jets with only one track, as is relevant for the search for one-prong 
hadronic tau decays, the optimisation 
again resulted in a value of $\mathcal{D}>0.95$. In this case
the signal efficiency was 
$\epsilon_{\rm{sig}}=22\pm0.2\%$, the background rejection was $R=637\pm41$ and the
separation was $S=5.5\pm0.2$.
These results are independent of the
model chosen for the simulation of the QCD cascade in the CC DIS simulation
(CDM or MEPS).

\subsection{Misidentification of QCD jets and electrons}
Both the suppression of QCD jets and the probability to misidentify electrons
as tau jets were determined from samples of simulated NC DIS events and
a selection of NC DIS data events \cite{thesis:moritz:2002}, where an electron
is scattered back-to-back to a jet in the detector.
The main selection criteria were $Q^2>400\gev^2$, where $Q^2$ is the virtuality of 
the exchanged boson, a well reconstructed electron
and at least one jet in the acceptance of the detector.

To determine the rejection factor for QCD jets, the electron-rejection
cuts from the CC DIS control selection described above were first applied
to all jets in the samples. For the surviving one-track jets, the tau discriminant
gives a further rejection factor of $R=550$. This result is in agreement
with the results from the CC DIS MC. No significant dependence 
of the rejection on the transverse energy of the jets was found. 
The results on the jet misidentification are the same in the data and in the simulation.

To determine the electron rejection, events in the NC DIS MC which had
no well-identified electron were also considered. The upper limit on
the fraction of NC DIS electrons that passed the tau selection and the CC
DIS control-selection cuts was $3\times 10^{-6}$. No difference between data
and simulation was observed.

\section{Event selection}
\label{sec-analysis}
The event selection closely follows the previous ZEUS search
 \cite{desy-03-012} for events
with isolated leptons and large missing transverse momentum.
The selection is based on the requirement of an isolated tau lepton,
decaying to one charged particle, together with large missing transverse momentum.
In a final selection stage, events with large values of the hadronic
transverse momentum were isolated.
Details of the analysis can be found elsewhere \cite{thesis:dannheim:2003}.
In the following, only the main selection criteria are described.

\subsection{Preselection of isolated tau events}
\begin{figure}[htbp]
\begin{minipage}[t]{1.0\linewidth}
\begin{center}
\epsfig{figure=plots/zeus.eps, width={0.13\linewidth},angle=0,clip=}
\end{center}
\end{minipage}
\begin{center} 
\epsfig{figure = 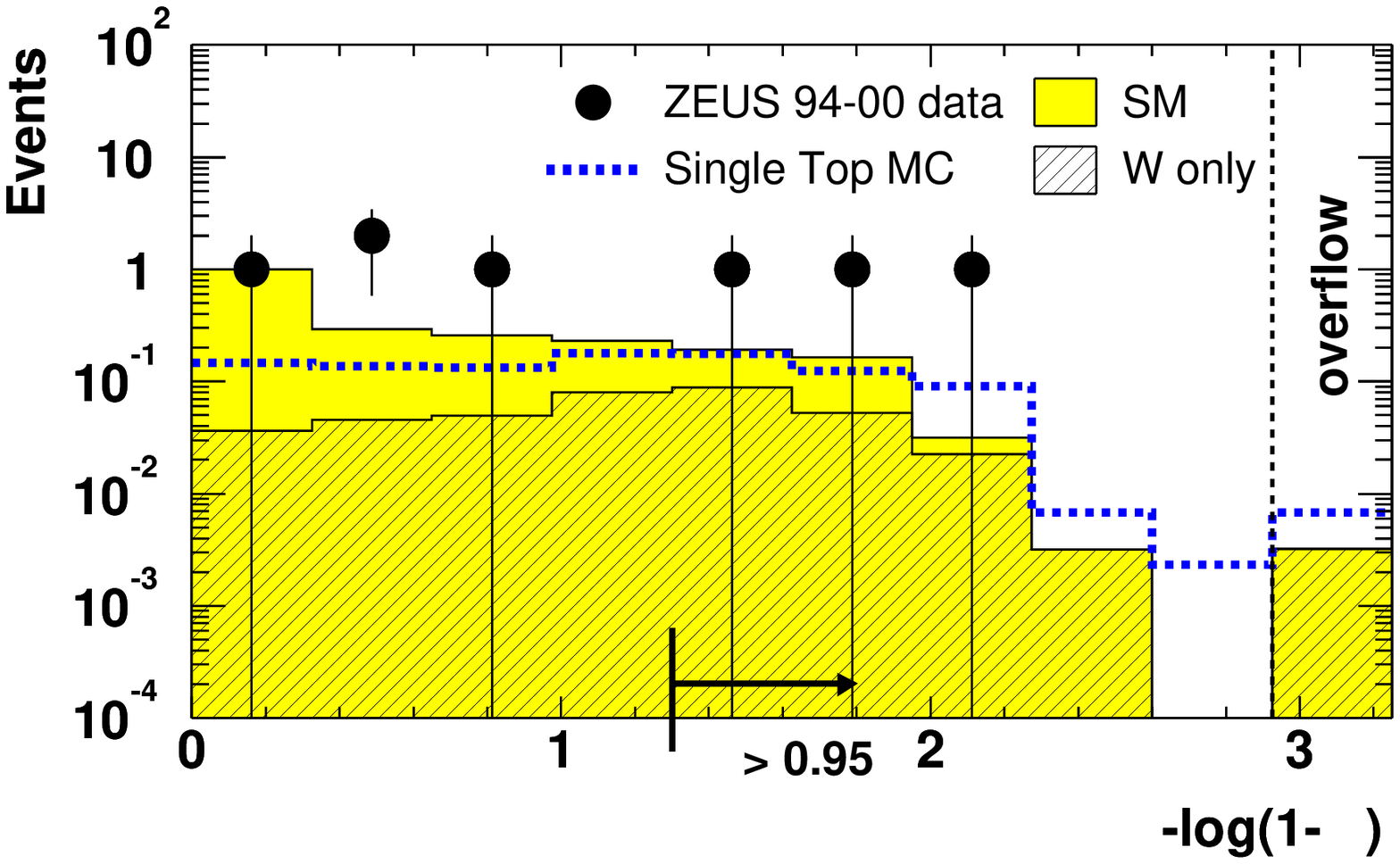,
width={13cm}, angle=0, clip=} \end{center}
\begin{center} \epsfig{figure = 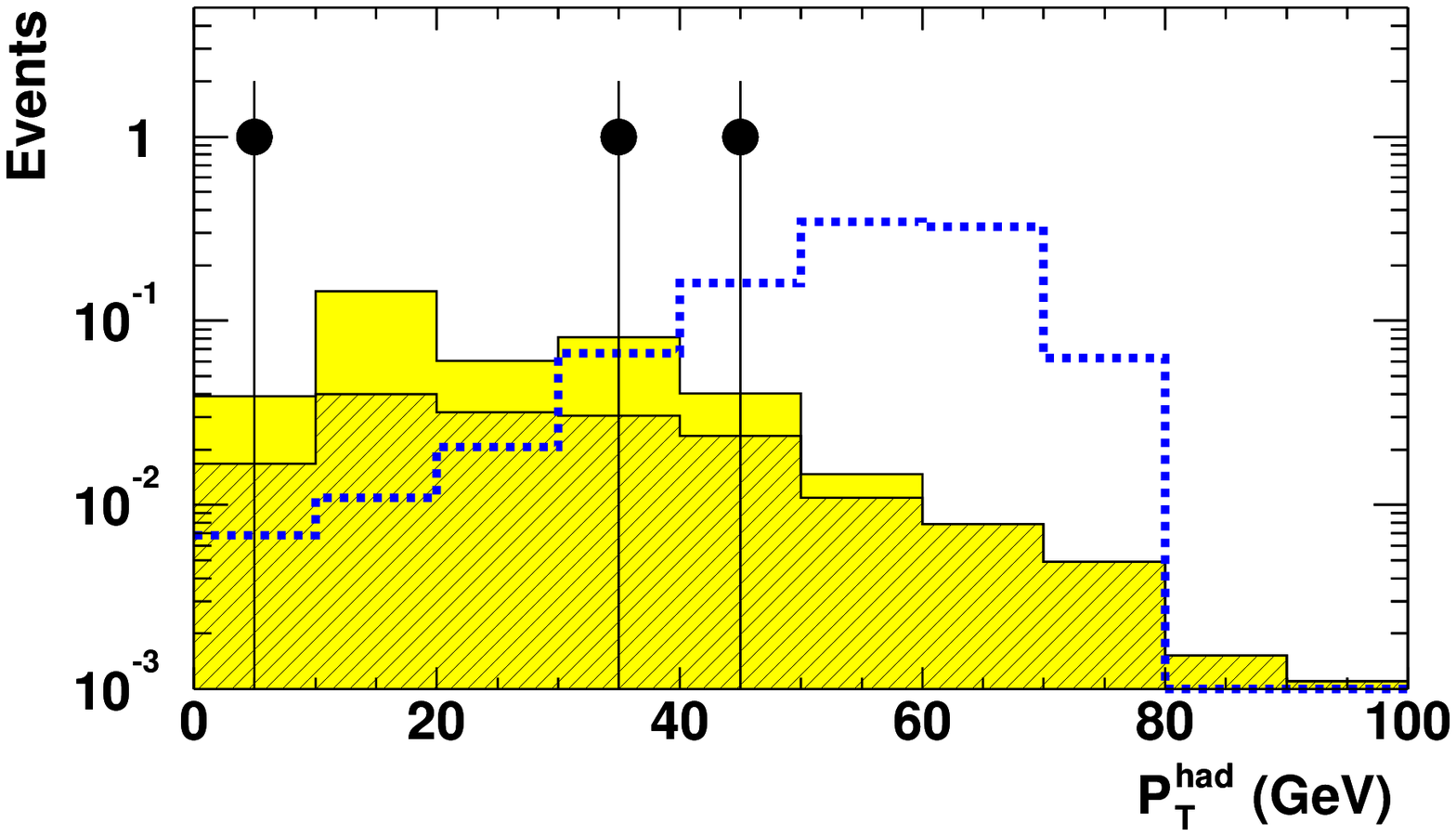,
width={13cm}, angle=0, clip=} \end{center}
\begin{picture} (0.,0.)
\setlength{\unitlength}{1.0cm}
\put (4.5,15.05){\bf\small (a)}
\put (4.5,7.15){\bf\small (b)}
\put (7.92,9.4){\large $\mathcal{D}$}
\put (13.23,8.80){\large $\mathcal{D}$}
\put (11.5,6.8){\large{$\mathcal{D}$}}
\put (11.98,6,835){$>$ \bf \textsf{0.95}}
\end{picture}
\caption{Distribution of (a) the tau discriminant, $-\log (1-\mathcal{D})$, for the tau
preselection before applying the cut $\mathcal{D}>0.95$ and (b)
the hadronic transverse momentum, $p_T^\mathrm{had}$, 
after applying the cut $\mathcal{D}>0.95$.
The data (points) are compared to the SM expectations 
(shaded histogram). The
hatched histogram represents the contribution from $W^\pm$ boson
production in the SM. The dashed line represents the distribution of
the single-top MC, including all decay channels of the $W^\pm$ boson,
normalised to an integral of one event.}
\label{fig-taupresel}
\end{figure}

A preselection of tau-candidate events was made as follows:

\begin{itemize}
\item 
   cuts on the CAL timing and $Z$ coordinate ($|Z|<50$~cm) of the
   event vertex along with algorithms based on the pattern of tracks in the
   CTD were used to reject events not originating from $e^\pm p$ collisions;
\item
   a large missing transverse momentum was
   required, $p_T^{\rm{CAL}}>20\gev$, where $p_T^{\rm{CAL}}$ was reconstructed
   using the energy deposited in the CAL cells, after corrections for
   non-uniformity and dead material located in front of the
   CAL~\cite{epj:c11:427};
\item the selected events had to contain at least one jet, reconstructed as
   described in Section \ref{sec-jetobs}, with a transverse energy 
   $E_T^{\rm{jet}}>5\gev$ within the range of $-1.0 < \eta < 2.5$;
\item a track with transverse
   momentum $p_T^{\rm{track}}>5\gev$, associated with the event vertex
   and pointing in the direction of a tau-candidate jet, was required. 
   It had to pass through at least three radial superlayers of the CTD (corresponding
   to $\theta \gtrsim 0.3\rad$) and to have $\theta<2\rad$.
   The track was required to be isolated with respect to all other tracks
   and jets in the event: $D_{\rm{trk}}>0.5$ and $D_{\rm{jet}}>1.8$, where
   $D_{\rm{trk}}$ and $D_{\rm{jet}}$ are the separation of the given track in the
   \{$\eta,\phi$\}-plane from the nearest neighbouring track and the
   nearest neighbouring jet in the event, respectively;
\item isolated tracks that were identified as electrons or muons\footnote{Electron candidates
   were identified using an algorithm that combined CAL and CTD information \cite{zfp:c74:207}.
   The identification of muons was based on the pattern of
   energy deposits in the CAL~\pcite{thesis:dannheim:2003}.} were rejected.
   An additional electron rejection was applied based on the fraction of electromagnetic jet energy
   and on the fraction of jet energy carried by the isolated track.
   Remaining electron-type events with a topology characteristic for NC DIS events 
   were rejected by requiring
   the acoplanarity, $\phi_\mathrm{acopl}^\mathrm{trk}$, to be greater than 
   $8^{\circ}$, where $\phi_\mathrm{acopl}^\mathrm{trk}$ is 
   defined as the azimuthal angle between the isolated track and the vector which balances
   the hadronic system. The four-vector of the hadronic system was calculated by subtracting
   the four-vector of the tau-candidate jet from the four-vector obtained from the
   energy deposited in the CAL cells.
   
\end{itemize}
   After this preselection,
   seven events remained, while $2.2^{+0.39}_{-0.58}$ are expected from SM
   background (18\% of the SM background came from single $W^\pm$ boson production).
   Table~\ref{tab-tauyields} summarises the event yields at
   different selection stages.
   The quoted uncertainties on the SM expectations are discussed in Section~\ref{sec-syst}.
   The discriminant distribution for these seven events is shown in Fig.~\ref{fig-taupresel}a
   as $-\log (1-\mathcal{D})$, to emphasise the high-discriminant region.

   Three out of the seven events have a tau discriminant
   $\mathcal{D}>0.95$ and are
   therefore likely to come from tau decay. 
   After applying the cut $\mathcal{D}>0.95$, $0.40^{+0.12}_{-0.13}$ events are
   expected from SM background (43\% from single $W^\pm$-boson production).
   Figure \ref{fig-taupresel}b shows the distribution of the transverse momentum
   of the hadronic system, $p_T^\mathrm{had}$, after applying the cut at
   $\mathcal{D}>0.95$. 

   The online event selection required significant missing transverse momentum 
   and a reconstructed vertex consistent with an $e^\pm p$ interaction. The efficiency
   of this online selection for the kinematic range of interest was found to 
   be 100\% for simulated events. 

\subsection{Selection of events with high hadronic transverse momentum}

\begin{figure}[htbp]
\begin{center} \epsfig{figure = 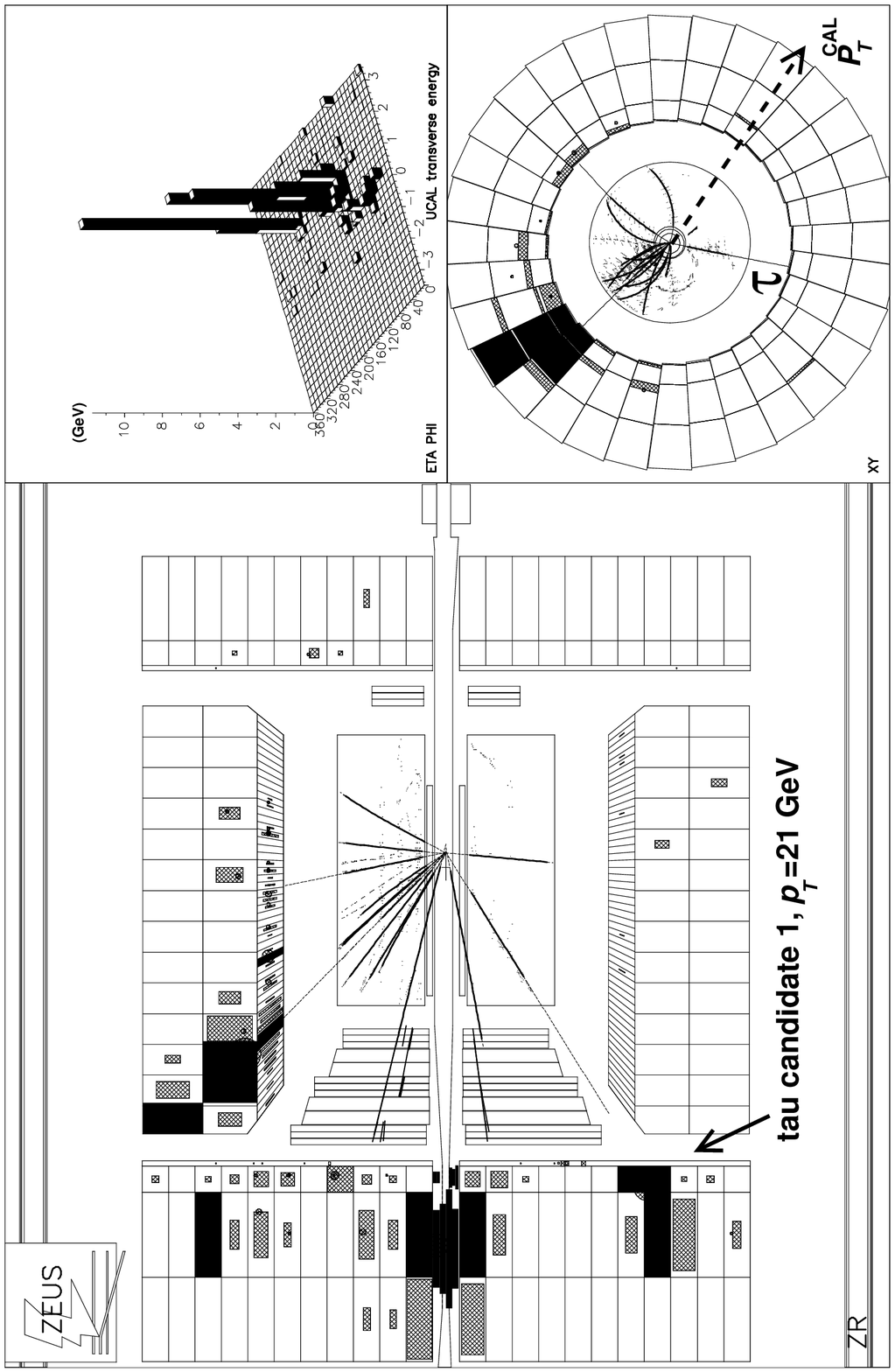,
width={9.5cm}, angle=-90, clip=} 
\end{center}
\begin{center} \epsfig{figure = 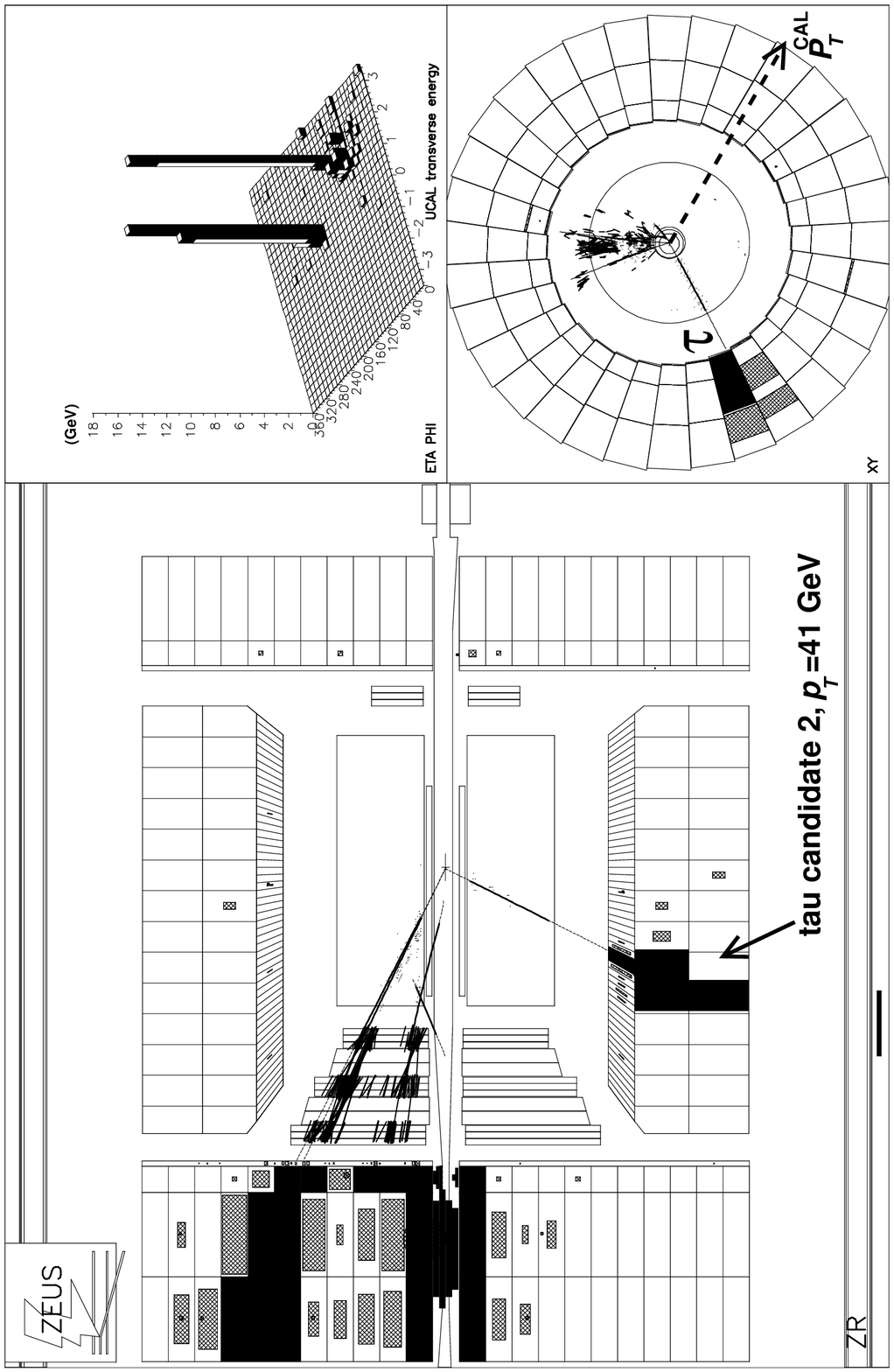,
width={9.5cm}, angle=-90, clip=} 
\end{center}
\caption{Tau-candidate events from $e^+p$ interactions at
$\sqrt{s}=318\gev$ in the ZEUS detector. 
The energy deposition in the CAL is proportional to the size and
density of shading in the CAL cells.
The lego plot shows the CAL energy deposition projected in the \{$\eta,\phi$\}-plane.
In the x-y-view, only the energy deposition in the barrel calorimeter
is shown.
The dashed arrow in the x-y view indicates
the direction of the missing transverse momentum in the calorimeter, $p_T^\mathrm{CAL}$.
Selected event variables for the two candidates are given in 
Table \ref{tab-kinematics}.
} 
\label{fig-events}
\end{figure}

   To design the final cut for events with high $p_T^\mathrm{had}$, the single-top
   MC was used as a template for the production and decay of a heavy state. 
   Following the published analysis in the electron and muon channels,
   an optimisation was performed, resulting in a cut at
   $p_T^\mathrm{had}>25\gev$, which gave the best separation between the single-top events 
   and the SM background. 
   Two events remained in the data, while $0.20\pm 0.05$ events are
   expected from the SM (49\% from single $W^\pm$-boson production). 
   With a higher cut at $p_T^\mathrm{had}>40\gev$, one event remains in the data,
   while $0.07\pm 0.02$ events are expected from the SM (71\% from
   single $W^\pm$-boson production). 
   Figure \ref{fig-events} shows event displays
   of the two events with large values of $p_T^\mathrm{had}$.
   Selected event variables for the two candidates are given in Table \ref{tab-kinematics}.
   Both events were found at large acoplanarity.
   The transverse mass was calculated from the tau-candidate jet and the missing
   transverse momentum as
   $M_T=\sqrt{2p ^\mathrm{jet} _T p_T^\mathrm{CAL} (1-\cos(\delta \phi_\mathrm{jet}))}$,
   where $\delta \phi_\mathrm{jet}$ is the angular difference in the 
   azimuthal plane between the tau jet and the direction of
   $p_T^\mathrm{CAL}$.  Both events were identified in the $e^+p$ data sample.

\subsection{Systematic uncertainties}
\label{sec-syst}
The errors on the background-expectation values were obtained as the quadratic
sum of the statistical uncertainties of the generated
MC events and each of the following systematic uncertainties:
\begin{itemize}
\item \emph{Simulation of the QCD cascade}.
The use of MEPS instead of CDM to estimate both the NC DIS and CC DIS background
gave a change of up to $-20\%$ in the total background estimation;
\item \emph{Track selection}.
A variation of the track-quality requirements and the angular range of the track
selection resulted in changes of up to $\pm15\%$ in the background estimation;
\item \emph{W cross section}.
The uncertainty for the expectation from single $W^\pm$-boson production,
after including higher-order QCD corrections by reweighting the LO MC 
samples~\cite{jp:g25:1434,*hep-ph/9905469,*hep-ph/0203269,*hep-ph/w_nlo_mc},
was estimated to be 15\%;
\item \emph{Tau-decay simulation}.
As a cross check, TAUOLA was used instead of JETSET for the simulation of the tau 
decays originating from single $W^\pm$-boson production. The TAUOLA program takes into account polarisation 
effects, whereas in JETSET the tau leptons are always decayed isotropically in their 
rest frame. The influence of
the tau-decay treatment on the jet-shape observables and on the efficiency for the
event selection was found to be negligible;
\item \emph{Tau-discriminant method}.
Both the CC DIS control selection and the tau-search
analysis were repeated with modified sets of jet-shape observables. In addition,
the box size used to evaluate the signal and background
densities was varied.
The dependence on these parameters was negligible;
\item \emph{Calorimeter energy scale}.
The uncertainty of $\pm 1\%$ on the absolute energy scale of both the electromagnetic
and the hadronic parts of the CAL resulted in changes
of up to $\pm4\%$ in the SM background estimation.
\end{itemize}

\section{Discussion of results}
\label{sec-results}
Table \ref{tab:finalselection} gives the result for the
final selection in the tau channel as well as the results of the previous 
search in the electron and muon channel \cite{desy-03-012} for two different
values of the cut on $p_T^\mathrm{had}$. In the electron (muon) channel, two
(five) events were observed for $p_T^\mathrm{had}>25\gev$, in good agreement
with the SM prediction. No event was observed in either channel for 
$p_T^\mathrm{had}>40\gev$. In combination with a search in the hadronic
decay channel of the $W^\pm$ boson, where no excess above the SM prediction was found,
a limit on the cross section for single-top 
production of $\sigma(ep\rightarrow etX,\sqrt{s}=318\gev)<0.225\pb$ at 95$\%$
C.L. was obtained \cite{desy-03-012}. For the tau channel,
two events were 
observed for $p_T^\mathrm{had}>25\gev$. Only hadronic tau decays were considered and
very restrictive selection cuts had to be applied to suppress the large background from 
electrons and quark- or gluon-induced jets. Therefore the selection efficiency for 
SM $W^\pm$ production is much smaller in the tau channel than in the electron and 
muon channels.

The Poisson probability to
observe two or more events when $0.20\pm 0.05$ events are expected is $1.8\%$,
where the uncertainty on the SM prediction was taken into account.
The observed events would correspond to a cross section for single-top production
that is much higher than the excluded cross section,
if the SM branching ratios for the top quark are assumed. 

In addition, single-top production 
produces positively charged leptons, and single anti-top production from 
protons is relatively suppressed by the parton densities.
Therefore the observed events are unlikely to be explained by the
hypothesis of single-top production.

$R_p$-violating SUSY models can explain
enhanced tau-production rates above the SM expectations. Moreover, if third-generation
sleptons are lighter than sleptons of the first and second generation, a corresponding
enhancement for electrons and muons could be strongly suppressed. In such models, the
stop quark can be directly produced at HERA via an $R_p$-violating Yukawa coupling
and subsequently decay through $R_p$-violating or gauge couplings. In particular, the
three-body gauge decay 
$\tilde{t}\rightarrow \tau \tilde{\nu_\tau}b,~\tilde{\tau} \nu_\tau b$
with the subsequent decays 
$\tilde{\tau}  \rightarrow \tau \tilde{\chi}^0$,
$\tilde{\nu_\tau} \rightarrow \nu_\tau \tilde{\chi}^0$
would produce a final state with the characteristics of the observed events:
a high-$p_T$ tau lepton at large acoplanarity angle, missing transverse momentum 
and large hadronic transverse momentum. However, in this case the
tau candidate has the same charge as the
incoming lepton beam, which is only the case for one of the two 
events surviving the cuts.

\section{Conclusion}
\label{sec-conclusion}
A search for events containing isolated tau leptons, 
large missing transverse momentum and large hadronic transverse momentum, 
produced in $e^\pm p$ collisions
at HERA, has been performed
using $130~\pb^{-1}$ of integrated luminosity. 
Such a signature could be produced within the framework of many 
theories beyond the Standard Model. The selection required
isolated tracks with associated pencil-like jets coming from hadronic 
tau decays. 
A multi-observable discrimination technique was used, exploiting
the internal jet structure to discriminate between hadronic tau decays and 
quark- or gluon-induced 
jets. 
Three isolated tau candidates were found, while $0.40^{+0.12}_{-0.13}$  
were expected from Standard Model processes, mainly from charged
current deep inelastic scattering and single $W^{\pm}$-boson production. 
A more restrictive selection was applied to isolate tau leptons in events
with large missing transverse momentum produced together with a hadronic final 
state with high transverse momentum,
as expected from the decay of a heavy particle.
Two candidate events with a transverse momentum of the hadronic system 
$p_T^\mathrm{had}>25\gev$
have been observed, while $0.20\pm0.05$ events were 
expected from Standard Model processes. The Poisson probability to
observe two or more events, assuming only SM contribution, is $1.8\%$,
so a statistical fluctuation cannot be excluded.
When considered together with previously published results in the electron
and muon channels,
the two candidates are unlikely
to originate from anomalous single-top production or any other process
where the tau lepton is produced through the decay of a $W^\pm$ boson. 

\section*{Acknowledgements}
We thank the DESY Directorate for their strong support and encouragement.
The remarkable achievements of the HERA machine group were essential for
the successful completion of this work and are greatly appreciated. We
are grateful for the support of the DESY computing and network services.
The design, construction and installation of the ZEUS detector have been
made possible owing to the ingenuity and effort of many people from DESY
and home institutes who are not listed as authors.
The NLO calculations for $W^\pm$ production
were provided by K.P.~Diener, C.~Schwanenberger and M.~Spira.
We would like to thank B.~Koblitz for providing his range-searching 
algorithm, which was used for the calculation of the tau discriminant.

\vfill\eject

{
\def\bibname{\Large\bf References}
\def\refname{\Large\bf References}
\pagestyle{plain}
\ifzeusbst
  \bibliographystyle{./BiBTeX/bst/l4z_default}
\fi
\ifzdrftbst
  \bibliographystyle{./BiBTeX/bst/l4z_draft}
\fi
\ifzbstepj
  \bibliographystyle{./BiBTeX/bst/l4z_epj}
\fi
\ifzbstnp
  \bibliographystyle{./BiBTeX/bst/l4z_np}
\fi
\ifzbstpl
  \bibliographystyle{./BiBTeX/bst/l4z_pl}
\fi
{\raggedright
\bibliography{./BiBTeX/user/syn.bib,%
              ./BiBTeX/user/mybib.bib,%
              ./BiBTeX/bib/l4z_articles.bib,%
              ./BiBTeX/bib/l4z_books.bib,%
              ./BiBTeX/bib/l4z_conferences.bib,%
              ./BiBTeX/bib/l4z_h1.bib,%
              ./BiBTeX/bib/l4z_misc.bib,%
              ./BiBTeX/bib/l4z_old.bib,%
              ./BiBTeX/bib/l4z_preprints.bib,%
              ./BiBTeX/bib/l4z_replaced.bib,%
              ./BiBTeX/bib/l4z_temporary.bib,%
              ./BiBTeX/bib/l4z_zeus.bib}}

\providecommand{\etal}{et al.\xspace}
\providecommand{\coll}{Coll.\xspace}
\catcode`\@=11
\def\@bibitem#1{%
\ifmc@bstsupport
  \mc@iftail{#1}%
    {;\newline\ignorespaces}%
    {\ifmc@first\else.\fi\orig@bibitem{#1}}
  \mc@firstfalse
\else
  \mc@iftail{#1}%
    {\ignorespaces}%
    {\orig@bibitem{#1}}%
\fi}%
\catcode`\@=12
\begin{mcbibliography}{10}

\bibitem{epj:c5:575}
H1 \coll, C.~Adloff \etal,
\newblock Eur.\ Phys.\ J.{} {\bf C~5},~575~(1998)\relax
\relax
\bibitem{desy-02-224}
H1 \coll, V. Andreev \etal,
\newblock Phys.\ Lett.{} {\bf B~561},~241~(2003)\relax
\relax
\bibitem{pl:b471:411}
ZEUS \coll, J.~Breitweg \etal,
\newblock Phys.\ Lett.{} {\bf B~471},~411~(2000)\relax
\relax
\bibitem{desy-03-012}
ZEUS \coll, S.~Chekanov \etal,
\newblock Phys.\ Lett.{} {\bf B~559},~153~(2003)\relax
\relax
\bibitem{np:b454:527}
T. Han, R.D. Peccei and X. Zhang,
\newblock Nucl.\ Phys.{} {\bf B~454},~527~(1995)\relax
\relax
\bibitem{pl:b426:393}
V.F. Obraztsov, S.R. Slabospitsky and O.P. Yushchenko,
\newblock Phys.\ Lett.{} {\bf B~426},~393~(1998)\relax
\relax
\bibitem{pr:d58:073008}
T. Han et al.,
\newblock Phys.\ Rev.{} {\bf B~426},~073008~(1998)\relax
\relax
\bibitem{pr:d60:074015}
T. Han and J.L. Hewett,
\newblock Phys.\ Rev.{} {\bf D~60},~074015~(1999)\relax
\relax
\bibitem{pl:b457:186}
H. Fritzsch and D. Holtmannspotter,
\newblock Phys.\ Lett.{} {\bf B~457},~186~(1999)\relax
\relax
\bibitem{np:b397:3}
J.~Butterworth and H.~Dreiner,
\newblock Nucl.\ Phys.{} {\bf B~397},~3~(1993)\relax
\relax
\bibitem{pl:b270:81}
T.~Kon and T.~Kobayashi,
\newblock Phys.\ Lett.{} {\bf B~270},~81~(1991)\relax
\relax
\bibitem{pr:d59:095009}
W.~Porod,
\newblock Phys.\ Rev.{} {\bf D~59},~095009~(1999)\relax
\relax
\bibitem{zeus:1993:bluebook}
ZEUS \coll, U.~Holm~(ed.),
\newblock {\em The {ZEUS} Detector}.
\newblock Status Report (unpublished), DESY (1993),
\newblock available on
  \texttt{http://www-zeus.desy.de/bluebook/bluebook.html}\relax
\relax
\bibitem{pl:b293:465}
ZEUS \coll, M.~Derrick \etal,
\newblock Phys.\ Lett.{} {\bf B~293},~465~(1992)\relax
\relax
\bibitem{nim:a279:290}
N.~Harnew \etal,
\newblock Nucl.\ Inst.\ Meth.{} {\bf A~279},~290~(1989)\relax
\relax
\bibitem{npps:b32:181}
B.~Foster \etal,
\newblock Nucl.\ Phys.\ Proc.\ Suppl.{} {\bf B~32},~181~(1993)\relax
\relax
\bibitem{nim:a338:254}
B.~Foster \etal,
\newblock Nucl.\ Inst.\ Meth.{} {\bf A~338},~254~(1994)\relax
\relax
\bibitem{nim:a309:77}
M.~Derrick \etal,
\newblock Nucl.\ Inst.\ Meth.{} {\bf A~309},~77~(1991)\relax
\relax
\bibitem{nim:a309:101}
A.~Andresen \etal,
\newblock Nucl.\ Inst.\ Meth.{} {\bf A~309},~101~(1991)\relax
\relax
\bibitem{nim:a321:356}
A.~Caldwell \etal,
\newblock Nucl.\ Inst.\ Meth.{} {\bf A~321},~356~(1992)\relax
\relax
\bibitem{nim:a336:23}
A.~Bernstein \etal,
\newblock Nucl.\ Inst.\ Meth.{} {\bf A~336},~23~(1993)\relax
\relax
\bibitem{desy-92-066}
J.~Andruszk\'ow \etal,
\newblock Preprint \mbox{DESY-92-066}, DESY, 1992\relax
\relax
\bibitem{zfp:c63:391}
ZEUS \coll, M.~Derrick \etal,
\newblock Z.\ Phys.{} {\bf C~63},~391~(1994)\relax
\relax
\bibitem{acpp:b32:2025}
J.~Andruszk\'ow \etal,
\newblock Acta Phys.\ Pol.{} {\bf B~32},~2025~(2001)\relax
\relax
\bibitem{proc:chep:1992:222}
W.H.~Smith, K.~Tokushuku and L.W.~Wiggers.
\newblock \emph{Proc. Computing in High-Energy Physics (CHEP), Annecy, France,
  1992}, C.~Verkerk and W.~Wojcik~(eds.), p.222. CERN, Geneva, Switzerland
  (1992). Also in preprint DESY-92-150B\relax
\relax
\bibitem{tech:cern-dd-ee-84-1}
R.~Brun et al.,
\newblock {\em {\sc geant3}},
\newblock Technical Report CERN-DD/EE/84-1, CERN, 1987\relax
\relax
\bibitem{np:b375:3}
U.~Baur, J.A.M.~Vermaseren and D.~Zeppenfeld,
\newblock Nucl.\ Phys.{} {\bf B~375},~3~(1992)\relax
\relax
\bibitem{cpc:39:347}
T.~Sj\"ostrand,
\newblock Comp.\ Phys.\ Comm.{} {\bf 39},~347~(1986)\relax
\relax
\bibitem{cpc:43:367}
T.~Sj\"ostrand and M.~Bengtsson,
\newblock Comp.\ Phys.\ Comm.{} {\bf 43},~367~(1987)\relax
\relax
\bibitem{cpc:82:74}
T.~Sj\"ostrand,
\newblock Comp.\ Phys.\ Comm.{} {\bf 82},~74~(1994)\relax
\relax
\bibitem{tauola:91}
S.~Jadach, Z.~Was and J.H.~Kuehn,
\newblock Comp.\ Phys.\ Comm.{} {\bf 64},~275~(1991)\relax
\relax
\bibitem{jp:g25:1434}
P.\ Nason, R.\ R\"uckl and M.\ Spira,
\newblock J.\ Phys.{} {\bf G~25},~1434~(1999)\relax
\relax
\bibitem{hep-ph/9905469}
M.~Spira,
\newblock Preprint \mbox{DESY-99-060} (\mbox{hep-ph/9905469}), 1999\relax
\relax
\bibitem{hep-ph/0203269}
K.P.~Diener, C.~Schwanenberger and M.~Spira,
\newblock Eur.\ Phys.\ J.{} {\bf C~25},~405~(2002)\relax
\relax
\bibitem{hep-ph/w_nlo_mc}
K.P.~Diener, C.~Schwanenberger and M.~Spira,
\newblock Preprint \mbox{hep-ex/0302040}, 2003\relax
\relax
\bibitem{pr:d55:1280}
H.L.~Lai \etal,
\newblock Phys.\ Rev.{} {\bf D~55},~1280~(1997)\relax
\relax
\bibitem{zfp:c56:589}
P.~Aurenche \etal,
\newblock Z.\ Phys.{} {\bf C~56},~589~(1992)\relax
\relax
\bibitem{cpc:81:381}
K.~Charchula, G.A.~Schuler and H.~Spiesberger,
\newblock Comp.\ Phys.\ Comm.{} {\bf 81},~381~(1994)\relax
\relax
\bibitem{cpc:69:155}
A.~Kwiatkowski, H.~Spiesberger and H.-J.~M\"ohring,
\newblock Comp.\ Phys.\ Comm.{} {\bf 69},~155~(1992)\relax
\relax
\bibitem{cpc:101:108}
G.~Ingelman, A.~Edin and J.~Rathsman,
\newblock Comp.\ Phys.\ Comm.{} {\bf 101},~108~(1997)\relax
\relax
\bibitem{cpc:71:15}
L.~L\"onnblad,
\newblock Comp.\ Phys.\ Comm.{} {\bf 71},~15~(1992)\relax
\relax
\bibitem{epj:c12:375}
CTEQ \coll, H.L.~Lai \etal,
\newblock Eur.\ Phys.\ J.{} {\bf C~12},~375~(2000)\relax
\relax
\bibitem{cpc:46:43}
M.~Bengtsson and T.~Sj\"ostrand,
\newblock Comp.\ Phys.\ Comm.{} {\bf 46},~43~(1987)\relax
\relax
\bibitem{cpc:136:126}
T.~Abe,
\newblock Comp.\ Phys.\ Comm.{} {\bf 136},~126~(2001)\relax
\relax
\bibitem{lsuhe-145-1993}
H.J.~Kim and S.~Kartik,
\newblock Preprint \mbox{{LSUHE}-145-1993}, 1993\relax
\relax
\bibitem{thesis:nguyen:2002}
C.N.~Nguyen.
\newblock Diploma Thesis, Univ. Hamburg, Hamburg (Germany), Report
  \mbox{DESY-THESIS-2002-024}, 2002\relax
\relax
\bibitem{np:b406:187}
S.~Catani \etal,
\newblock Nucl.\ Phys.{} {\bf B~406},~187~(1993)\relax
\relax
\bibitem{pr:d48:3160a}
S.D.~Ellis and D.E.~Soper,
\newblock Phys.\ Rev.{} {\bf D~48},~3160~(1993)\relax
\relax
\bibitem{proc:snowmass:1990:134}
J.E.~Huth \etal,
\newblock {\em Research Directions for the Decade. Proceedings of Summer Study
  on High Energy Physics, 1990}, E.L.~Berger~(ed.), p.~134.
\newblock World Scientific (1992).
\newblock Also in preprint \mbox{FERMILAB-CONF-90-249-E}\relax
\relax
\bibitem{pl:b531:9}
ZEUS \coll, S.~Chekanov \etal,
\newblock Phys.\ Lett.{} {\bf B~531},~9~(2002)\relax
\relax
\bibitem{np:b545:3}
H1 \coll, C.~Adloff \etal,
\newblock Nucl.\ Phys.{} {\bf B~545},~3~(1999)\relax
\relax
\bibitem{desy-02-217}
ZEUS \coll, S.~Chekanov \etal,
\newblock Phys.\ Lett.{} {\bf B~558},~41~(2003)\relax
\relax
\bibitem{jhep:09:009}
J.R. Forshaw and M.H. Seymour,
\newblock JHEP{} {\bf 9909},~009~(1999)\relax
\relax
\bibitem{np:b421:545}
M.H. Seymour,
\newblock Nucl.\ Phys.{} {\bf B~421},~545~(1994)\relax
\relax
\bibitem{thesis:dannheim:2003}
D. Dannheim.
\newblock PhD Thesis, Univ. Hamburg, Hamburg (Germany), Report
  \mbox{DESY-THESIS-2003-025}, 2003\relax
\relax
\bibitem{pr:d65:092004}
ZEUS \coll, S. Chekanov \etal,
\newblock Phys.\ Rev.{} {\bf D~65},~092004~(2002)\relax
\relax
\bibitem{carli:modt:0011224}
T. Carli and B. Koblitz,
\newblock {\em Proceedings of the {VII} International Workshop on Advanced
  Computing and Analysis Techniques in Physics Research}, P.~Bhat and
  M.~Kasemann~(eds.), p.~110.
\newblock American Institute of Physics (2000).
\newblock Also in hep-ph/0011224\relax
\relax
\bibitem{carli:modt:0211019}
T. Carli and B. Koblitz,
\newblock Nucl.\ Inst.\ Meth.{} {\bf A~501},~576~(2003)\relax
\relax
\bibitem{thesis:moritz:2002}
M.~Moritz.
\newblock Ph.D.\ Thesis, Univ. Hamburg, Hamburg (Germany), Report
  \mbox{DESY-THESIS-2002-009}, 2002\relax
\relax
\bibitem{epj:c11:427}
ZEUS \coll, J.~Breitweg \etal,
\newblock Eur.\ Phys.\ J.{} {\bf C~11},~427~(1999)\relax
\relax
\bibitem{zfp:c74:207}
ZEUS \coll, J.~Breitweg \etal,
\newblock Z.\ Phys.{} {\bf C~74},~207~(1997)\relax
\relax
\end{mcbibliography}
}
\vfill\eject

%
%
%
\newpage

\begin{table}
\begin{center}
\begin{tabular}{|l||l|l|l|c|} \hline
Selection stage & Obs. & SM exp. ($W^\pm$ contrib.) & $\epsilon _\mathrm{s.top} \cdot BR$ (\%)   \\
\hline \hline
Isolated tracks & 7 & 2.18$^{+0.39}_{-0.58}$~ (18\%) & 0.68 \\
\hline
Discriminant $\mathcal{D}>0.95$ & 3 & 0.40$^{+0.12}_{-0.13}$~ (43\%)  & 0.27 \\ 
\hline
$p_T^\mathrm{had}>25\gev$ (final sel.)& 2 & 0.20$^{+0.05}_{-0.05}$~ (49\%) & 0.27 \\
\hline
$p_T^\mathrm{had}>40\gev$ & 1 & 0.07$^{+0.02}_{-0.02}$~ (71\%) & 0.25 \\
\hline
\end{tabular}
\end{center}
\caption[~Event yields tau search]{Event yields for the data from 1994-2000, corresponding background 
expectations and efficiency times branching ratio for the single-top MC
at different selection stages in the search for isolated tau leptons. The percentage
of single-$W$ production included in the expectation is indicated in parentheses. The statistical
and systematic uncertainties in quadrature are also indicated.}
\label{tab-tauyields}
\end{table}

\begin{table}
\begin{center}
\begin{tabular}{|l||l|l|} \hline
Quantity & Event 1 & Event 2 \\
\hline \hline
Missing transverse momentum $p_T^\mathrm{CAL}$ & 37~GeV & 39~GeV \\
\hline
Hadronic transverse momentum $p_T^\mathrm{had}$        & 48~GeV & 38~GeV \\
\hline
Transverse momentum of the tau-candidate jet $p_T^\mathrm{jet}$        & 21~GeV & 41~GeV \\
\hline
Transverse momentum of the tau-candidate track $p_T^\mathrm{trk}$        & 9~GeV & 27~GeV \\
\hline
Charge sign of the tau-candidate track   & $-$  & $+$ \\
Significance in numbers of standard deviations & $5.7\sigma$  & $3.8\sigma$  \\
\hline
Acoplanarity of the tau-candidate track $\phi_\mathrm{acopl}^\mathrm{trk}$  & $45^\circ$ & $55^\circ$ \\
\hline
Transverse mass $M_T$                   & 32~GeV & 70~GeV \\
\hline
Discriminant $\mathcal{D}$                                       & 0.994  & 0.977 \\
\hline
\end{tabular}
\end{center}
\caption{Selected event variables for the two tau-candidate events at high $p_T^\mathrm{had}$.}
\label{tab-kinematics}
\end{table}

\begin{table}
\begin{center}
\begin{tabular}{|c|c|c|c|} \hline
ZEUS  & Electron & Muon & Tau \\
1994-2000 $e^\pm p$ & obs./exp. & obs./exp. & obs./exp.  \\
$\Lumi = 130.1\pbi$ & ($W^\pm$ contribution) & ($W^\pm$ contribution) &  ($W^\pm$ contribution) \\ \hline 
$p_T^\mathrm{had}~>~25\gev$ &  2 / $2.90~^{+0.59}_{-0.32}~(45\%)$   & 5 / 2.75~$^{+0.21}_{-0.21}$ (50\%)  & 2 / 0.20~$^{+0.05}_{-0.05}$~(49\%) \\ \hline
$p_T^\mathrm{had}~>~40\gev$ &  0 / $0.94~^{+0.11}_{-0.10}~(61\%)$   & 0 / $0.95~^{+0.14}_{-0.10}~(61\%)$  & 1 / 0.07~$^{+0.02}_{-0.02}$~(71\%) \\ \hline
\end{tabular}
\end{center}
\caption{Summary of the results of searches for events with isolated
leptons, missing transverse momentum and large $p_T^\mathrm{had}$. The number
of observed events is compared to the SM prediction. The $W^\pm$
component is given in parentheses in percent. The statistical and systematic uncertainties added
in quadrature are also indicated.
The results for the electron and the muon channel were
obtained from a previous search \pcite{desy-03-012}.}
\label{tab:finalselection}
\end{table}

\newpage

\vfill\eject

%
%
\end{document}